\documentclass[aps,prresearch,twocolumn,superscriptaddress]{revtex4-2}

\usepackage{graphicx}
\usepackage{amsmath}
\usepackage{bm}
\usepackage{xcolor}

\begin{document}

\title{Ponderomotive Achromat for Electron Optics: Radially Polarized Annular Focusing and a Round-Lens Corrector Regime}

\author{Yuuki Uesugi}
    \email{yuuki.uesugi@aist.go.jp}
    \affiliation{Sensing Technology Research Institute, National Institute of Advanced Industrial Science and Technology, 807-1 Syukumachi, Tosu, Saga 841-0052, Japan}
\author{Yuichi Kozawa}
    \affiliation{Institute of Multidisciplinary Research for Advanced Materials, Tohoku University, 2-1-1 Katahira, Aoba-ku, Sendai, Miyagi 980-8577, Japan}

\date{April 13, 2026; revised on July 23, 2026; accepted on August 24, 2026}

\begin{abstract}
Ponderomotive electron optics has attracted attention because structured optical fields provide round-lens functionalities, including negative lens power and negative spherical aberration, that are difficult to achieve with conventional Einzel and solenoid lenses.
How ponderomotive lenses disperse with electron energy, and whether that dispersion can be engineered for achromatization, have remained largely unexplored.
Here we show that the transverse and longitudinal components of a vector optical field exhibit distinct energy dispersions because a relativistic correction modifies the effective lens action of the component polarized parallel to the electron velocity.
In a radially polarized Bessel field coaxial with the electron beam, the two components form spatially co-located $J_1^2$- and $J_0^2$-type ponderomotive potentials whose relative weights depend on the electron energy, the Bessel cone angle, and the propagation direction of the light relative to the electrons.
Using a local Abbe number defined in the energy domain, we find that this single field can operate as a self-achromatic round lens.
We further construct zero-separation doublets---with a magnetic objective-lens model or an azimuthally polarized Bessel lens---that simultaneously cancel the first-order axial chromatic- and third-order spherical-aberration coefficients at a fixed electron-optical focal length, with the useful solutions occurring for counter-propagating light.
Finite-aperture calculations at optical wavelengths of $1064~\mathrm{nm}$ and $10.6~\mathrm{\mu m}$ relate the achievable probe size to the required optical power, and at $10.6~\mathrm{\mu m}$ they identify a low-energy, large-energy-spread regime in which the corrected doublets outperform the magnetic objective-lens model.
The required optical power is compatible with demonstrated near-infrared ultrashort-pulse sources, so a single-pass proof-of-principle experiment is within reach.
\end{abstract}

\maketitle

\section{Introduction}

Chromatic aberration remains one of the primary constraints in electron optics.
For a fixed electron-lens architecture, an energy spread in the electron beam is converted into a focal shift; this focal shift produces a resolution-degrading disk in the image plane in imaging optics, or increases the probe size and thereby degrades spatial resolution in scanning microscopy \cite{Scherzer1936, HawkesKasper1989, Rose2009, Hawkes2015}.
Chromatic-aberration correctors exist, but they are realized as electromagnetic multipole assemblies, which makes them bulky, alignment-sensitive, and strongly coupled to the rest of the column \cite{Haider2009, Linck2016}.

In practical high-resolution electron microscopy, the standard route to mitigating chromatic aberration is to reduce the energy spread before the objective lens.
Modern high-performance instruments therefore commonly employ electron monochromators \cite{Kimoto2014, Krivanek2019}.
Source-level developments are also aimed at electron beams with energy spreads below those available from conventional cold-field-emission cathodes, which have long served as the classical high-brightness source for electron microscopy.
Representative directions include ultracold electron sources and highly monochromatic planar emitters \cite{Engelen2014, deRaadt2023, Morishita2020, Murakami2020, Igari2021}.

Monochromatization is an effective strategy for microscopy aimed at ultimate spatial or spectroscopic resolution.
However, there remain electron-optical systems in which a broad energy spread is intrinsic or operationally difficult to avoid, and in such cases chromatic-aberration correction remains a central requirement.
Examples include ultrahigh-voltage electron microscopy based on radio-frequency or linear acceleration schemes, and attosecond electron-bunch compression based on energy modulation and temporal-lens effects, where energy spreads can reach the order of 100 eV \cite{Hilbert2009, Morimoto2018, Kozak2018, Verhoeven2018, Sannomiya2019}.
Focusing such beams down to nanometer-scale probe sizes is not yet a routine capability.

This motivates the search for compact alternatives, including optical-field approaches in which an electron lens is generated via the ponderomotive interaction with a strong optical field \cite{Minogin2009, Uesugi2021, Uesugi2022, Mihaila2022, Uesugi2023, Rollo2026}.
In a classical description, an electron in an inhomogeneous oscillating optical field undergoes a rapid quiver motion and experiences a cycle-averaged ponderomotive force \cite{Kibble1966, Startsev1997}.
This force can be represented by an effective interaction potential, the ponderomotive potential, which is proportional to the local optical intensity.
For an electron wave function, this light-induced potential acts as an electron-optical phase-modulating potential, in analogy with the role played by electrostatic and magnetostatic fields in conventional electron optics.

We refer to electron lenses generated by a ponderomotive potential as ponderomotive lenses.
A useful feature of a ponderomotive lens is that its ``material'' is an optical field.
With modern laser beam-shaping and vector-beam techniques, the wavefront and polarization distribution of the optical beam can be spatially engineered.
Consequently, the potential landscape in the interaction region can be designed more flexibly than is possible with electrostatic or magnetic lenses.
Ponderomotive lenses can realize lens behaviors, including negative focal length and negative spherical aberration, that are difficult to obtain with conventional lenses such as Einzel and solenoid lenses.
Because spherical-aberration correction is central to atomic and sub-\AA{} electron microscopy \cite{Batson2002}, recent work has explored optical-field-based alternatives to complex multipole correctors, including near-field and free-space schemes \cite{Konecna2020, GarciadeAbajo2021, Mihaila2025OE, Uesugi2025, Mihaila2025NP, Nekula2025}.

Achromatization requires control of both the lens power and its dispersion with respect to the electron-beam energy.
In ordinary optics, this is achieved by combining positive and negative lens powers with different dispersions, so that the first derivative of the total power with respect to wavelength vanishes at the design wavelength \cite{BornWolf}.
Ponderomotive lenses already provide the required sign control of the lens power, while their dispersion has usually been regarded as essentially analogous to that of magnetic lenses, as discussed below.
Yet an experiment by Axelrod \textit{et al.} has already shown that the ponderomotive potential acquires an additional relativistic correction when the optical-field polarization is parallel to the electron-beam axis \cite{Axelrod2020}.
This result has so far not been exploited in lens design.
It implies that the dispersion of a ponderomotive lens can be engineered through the polarization structure: the longitudinal and transverse optical-field components contribute different energy dispersions, providing the dispersion contrast required for achromatic lens design.

Here we analyze the lens power, energy dispersion, and geometrical aberrations of ponderomotive round lenses formed by vector Bessel fields.
An azimuthally polarized Bessel field carries a purely transverse optical-field component and produces a ponderomotive potential with a $J_1^2$ profile.
A radially polarized Bessel field carries both transverse and longitudinal optical-field components, the latter polarized parallel to the electron-beam axis, and its ponderomotive potential is a weighted sum of $J_1^2$ and $J_0^2$ profiles.
Because the relativistic correction modifies the effective electron-lens action of the longitudinal contribution, the relative weights of the two contributions depend on the electron-beam energy, the Bessel cone angle, and the relative propagation direction of the electron and optical beams.
The two contributions thus provide the required dispersion contrast within a single field, and the resulting radial Bessel lens can exhibit either converging or diverging power, together with negative, zero, or positive chromatic dispersion.

\begin{figure}[t]
    \centering
    \includegraphics*[width=\columnwidth]{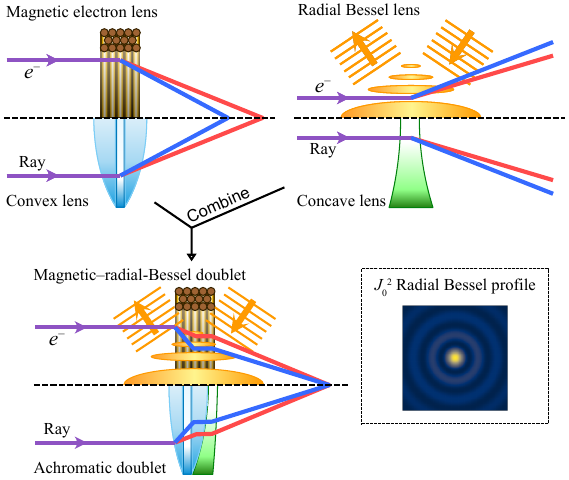}
\caption{
Conceptual illustration of achromatization by combining a magnetic electron lens with a radial Bessel lens.
The upper panels show a converging magnetic electron lens and a diverging radial Bessel lens; in each panel, the electron-lens action above the axis is paired with the corresponding refractive-lens analogue below.
The red and blue lines represent electrons of different energies in the upper half of each panel and optical rays of different wavelengths in the lower half, illustrating the chromatic dispersion of each lens.
When the two lenses are combined with zero separation, their dispersions cancel, and rays of different energies converge to a common point on the optical axis.
The lower-right panel shows the $J_0^2$ intensity profile used as the reference for the diverging radial Bessel lens.
}
    \label{fig:illustration}
\end{figure}

To characterize this range of dispersion behavior, we introduce an energy-domain local Abbe number and derive the corresponding chromatic-aberration coefficient.
The radial Bessel lens can become self-achromatic at a suitable cone angle.
We then combine it with either a magnetic objective-lens model or an azimuthal Bessel lens (Fig.~\ref{fig:illustration}).
For a fixed total electron-optical focal length, we identify zero-separation doublets that simultaneously cancel the first-order axial chromatic and third-order spherical aberrations.
These doublets follow the classic achromat strategy of combining lens powers with different dispersions, with the contrast supplied by the polarization-dependent ponderomotive response.

We further evaluate the finite-aperture performance of the singlets and doublets at optical wavelengths of $1064~\mathrm{nm}$ and $10.6~\mathrm{\mu m}$, using the third- and fifth-order spherical-aberration coefficients as well as the full spherical aberration computed from the Bessel profile.
This comparison shows that cancellation of the first-order chromatic- and third-order spherical-aberration coefficients does not by itself guarantee a useful electron objective lens, because fifth- and higher-order aberrations can remain large.
We relate the required integrated ponderomotive strength to the optical power of a finite Bessel-like field.
Finally, we discuss applications to electron beams whose broad or discrete energy distributions must be preserved, and outline extensions of the framework to compound electron-optical systems with additional dispersive and geometrical degrees of freedom.

\section{Methods}

\subsection{Energy-domain local Abbe number and achromat condition}

\subsubsection{Local dispersion and zero-separation achromat condition}

In ordinary optics, material dispersion is often summarized by the discrete Abbe number defined from refractive indices at standard Fraunhofer wavelengths \cite{BornWolf},
\begin{gather}
    V_\mathrm{d} = \frac{n_\mathrm{d}-1}{n_\mathrm{F}-n_\mathrm{C}}.
    \label{eq:Vd}
\end{gather}
Here $n_\mathrm{d}$, $n_\mathrm{F}$, and $n_\mathrm{C}$ are the refractive indices at the helium $d$ line ($\lambda_\mathrm{d}=587.56\,\mathrm{nm}$), the hydrogen $F$ line ($\lambda_\mathrm{F}=486.13\,\mathrm{nm}$), and the hydrogen $C$ line ($\lambda_\mathrm{C}=656.27\,\mathrm{nm}$), respectively.

For an ordinary refractive thin lens, the lens power is defined as $F=1/f$, where $f$ is the focal length.
It is determined by the product of the refractive-index contrast and a purely geometrical factor.
When the lens geometry is fixed, the wavelength dependence of the power therefore originates only from the material dispersion, and the power can be written as
\begin{gather}
    F(\lambda) = [n(\lambda)-1]\,S,
\end{gather}
where $n(\lambda)$ is the refractive index and $S$ is the shape factor determined by the surface curvatures.
This motivates the definition of a local continuous Abbe number directly in terms of the wavelength dependence of the power:
\begin{gather}
    \nu(\lambda) = -\frac{F(\lambda)}{\lambda \, (dF/d\lambda)}.
    \label{eq:abbe_lambda}
\end{gather}

For electron optics, we rewrite this quantity in terms of electron-beam energy.
Throughout this article, $\gamma$ denotes the electron Lorentz factor, $\beta=\sqrt{1-1/\gamma^2}$, and $v=\beta c$, where $c$ is the speed of light and $m$ is the electron rest mass.
The electron de Broglie wavelength is
\begin{gather}
    \lambda_e = \frac{h}{\beta\gamma mc},
    \label{eq:lambdae}
\end{gather}
with $h$ Planck's constant.
Since $\lambda_e \propto (\gamma^2-1)^{-1/2}$, one has
\begin{gather}
    \frac{d\lambda_e}{d\gamma} = -\lambda_e\,\frac{\gamma}{\gamma^2-1},
    \label{eq:dlamdg}
\end{gather}
and Eq.~(\ref{eq:abbe_lambda}) becomes
\begin{gather}
    \nu(\gamma) = \frac{\gamma}{\gamma^2 - 1} \frac{F(\gamma)}{dF/d\gamma},
    \label{eq:abbe}
\end{gather}
which serves as a local Abbe number in the energy domain.
Equivalently,
\begin{gather}
    \frac{dF}{d\gamma} = \frac{\gamma}{\gamma^2-1}\,\frac{F(\gamma)}{\nu(\gamma)}.
    \label{eq:dFdgamma}
\end{gather}

When two or more thin-lens components are co-located so that their effective separation is negligible, the system can be treated as a zero-separation lens system.
In that case, the total lens power is
\begin{gather}
    F_\mathrm{tot}(\gamma) = \sum_i F_i(\gamma).
    \label{eq:Ftot}
\end{gather}
Because this relation is linear in the lens power, the series expansion of $F_\mathrm{tot}$ around the design Lorentz factor $\gamma_0$ is obtained by summing the corresponding expansion terms of the individual powers.
The zeroth-order term gives the design power, whereas the first-order term describes the chromatic variation of the focusing strength, i.e., the first-order axial chromatic aberration.
Thus, cancelling this first-order term gives the condition for first-order achromatization.

The system is first-order achromatic at the design Lorentz factor $\gamma_0$ when
\begin{gather}
    \left.\frac{dF_\mathrm{tot}}{d\gamma}\right|_{\gamma=\gamma_0}=0.
\end{gather}
This condition can be rewritten in terms of the local Abbe numbers by substituting Eq.~(\ref{eq:dFdgamma}).
Since the factor $\gamma/(\gamma^2-1)$ is common to all lens elements at $\gamma_0$, it cancels out in the achromatization condition. Thus, the achromat condition becomes
\begin{gather}
    \sum_i \frac{F_i(\gamma_0)}{\nu_i(\gamma_0)} = 0.
    \label{eq:ach}
\end{gather}
The same first-order dispersion can also be expressed in terms of the axial chromatic-aberration coefficient.
For a parallel incident beam, we define
\begin{gather}
    \Delta f = C_\mathrm{c}\frac{\Delta T}{T},
    \label{eq:firstCc}
\end{gather}
where $T=(\gamma-1)mc^2$ is the electron kinetic energy.
Rewriting Eq.~(\ref{eq:firstCc}) in terms of $\gamma$ and using Eq.~(\ref{eq:abbe}) gives
\begin{gather}
    C_\mathrm{c} = (\gamma - 1)\frac{df}{d\gamma}
    = -\frac{\gamma - 1}{F^2}\frac{dF}{d\gamma}
    = -\frac{\gamma}{(\gamma + 1)F\nu}.
    \label{eq:Cc_dFdgamma}
\end{gather}
For a thin-lens system with zero separation and finite total lens power, the condition $dF_\mathrm{tot}/d\gamma=0$ is equivalent to $C_{\mathrm{c,tot}}=0$.

\subsubsection{Reference dispersion of conventional electron lenses}

The local Abbe number provides a common measure for comparing the energy dispersion of different electron-lens mechanisms.
As a reference, we first consider a magnetostatic solenoid lens with fixed geometry and excitation.
In the paraxial thin-lens approximation, its lens power scales with the inverse square of the electron momentum,
\begin{gather}
F_\mathrm{M}(\gamma)
=
\frac{\kappa_\mathrm{M}}{\beta^2\gamma^2}
=
\frac{\kappa_\mathrm{M}}{\gamma^2-1},
\label{eq:FM}
\end{gather}
where
\begin{gather}
\kappa_\mathrm{M}
=
\frac{e^2}{4m^2c^2}
\int_{-\infty}^{\infty} B_z^2(z) dz
\end{gather}
is independent of the electron energy for a fixed axial magnetic-field distribution $B_z(z)$.
Substitution of Eq.~(\ref{eq:FM}) into Eq.~(\ref{eq:abbe}) gives
\begin{gather}
\nu_\mathrm{M}
=
-\frac{1}{2}.
\label{eq:abbe_M}
\end{gather}
Thus, the magnetostatic lens constitutes a simple reference dispersion class whose local Abbe number is independent of the electron energy.

In an electrostatic round lens, the electron energy changes inside the lens field, and a quantitative treatment requires the energy-dependent trajectories for a specified electrode configuration.
For conceptual comparison, a weak electrostatic round lens may be described by a fixed-field, straight-trajectory model,
\begin{gather}
F_\mathrm{E}(\gamma)
=
\frac{\kappa_\mathrm{E}}{\beta^2\gamma}
=
\frac{\kappa_\mathrm{E}\gamma}{\gamma^2-1},
\label{eq:FE}
\end{gather}
where $\kappa_\mathrm{E}$ represents the electrode geometry and applied-potential distribution within this approximation; the factor of $\gamma$ in the numerator reflects the energy change of the electron inside the field.
Within this model, the local Abbe number is
\begin{gather}
\nu_\mathrm{E}(\gamma)
=
-\frac{\gamma^2}{\gamma^2+1},
\label{eq:abbe_E}
\end{gather}
illustrating that electrostatic and magnetostatic lenses can exhibit different relativistic dispersion even when both operate as conventional positive-power round lenses.

\subsection{Relativistic ponderomotive potential of vector Bessel fields}

As a rotationally symmetric structured optical field, we consider a Bessel-type optical beam \cite{Durnin1987, Herman1991} whose optical axis coincides with the electron-beam axis $z$.
Near the axis, the transverse structure is described by Bessel functions $J_n(Kr)$, where $J_n$ is the $n$th-order Bessel function of the first kind, $k$ is the optical wavenumber, and
\begin{gather}
    K = k\sin\theta
\end{gather}
is the transverse wavenumber for the Bessel cone half-angle $\theta$.
In vacuum, $\sin\theta$ is equal to the numerical aperture (NA) of the focusing optics used to form the Bessel field.
We therefore introduce
\begin{gather}
    s_\theta \equiv \sin\theta = \frac{K}{k} = \mathrm{NA},
\end{gather}
which is used as the geometrical cone parameter below.

Axelrod \textit{et al.} showed that the ponderomotive phase shift of a relativistic electron depends not only on the cycle-averaged optical intensity, but also on the electron velocity and the optical-field polarization \cite{Axelrod2020}.
Their formulation is useful here because the second-order interaction can be written in terms of the laboratory-frame Coulomb-gauge vector potential as
\begin{gather}
U = \frac{e^2}{2\gamma m}\left\langle \left(\mathbf{A}-\nabla G\right)^2 - \beta^2\left(A_z-\partial_zG\right)^2 \right\rangle ,
\label{eq:U_rel_general}
\end{gather}
where $e$ is the elementary charge, $\mathbf{A}=(A_r,A_\phi,A_z)$ is the optical vector potential, and $\langle\cdots\rangle$ denotes averaging over the optical cycle.
The gauge function $G$ is defined by
\begin{gather}
\left(\partial_t+\beta c\partial_z\right)G = \beta c A_z .
\label{eq:G_equation_general}
\end{gather}
Thus the gauge correction is relevant only when the optical field has a longitudinal vector-potential component $A_z$.
The original analysis of Axelrod \textit{et al.} treated a one-dimensional optical standing wave whose wavevector is perpendicular to the electron-beam direction.
Here we apply the same phase-based formulation to Bessel fields coaxial with the electron beam.
A key difference is that each plane-wave constituent of a propagating Bessel field has a finite longitudinal wavenumber $k\cos\theta$.

We consider two cylindrically symmetric vector Bessel fields generated by focusing azimuthally and radially polarized annular beams.
Such fields naturally arise in high-NA focusing of cylindrically polarized optical beams \cite{Richards1959,Quabis2000,Youngworth2000,Dorn2003,Sheppard2004,Kozawa2006}.
The azimuthally polarized field contains only the transverse azimuthal component $A_\phi$.
Its focal-plane intensity has a $J_1^2$-type profile, which vanishes on the optical axis and forms a concentric annular distribution.

\begin{figure}[t]
    \centering
    \includegraphics*[width=\columnwidth]{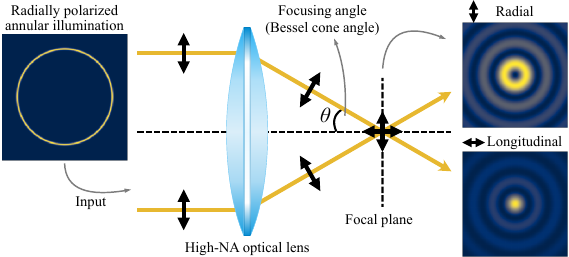}
    \caption{Formation of a radially polarized Bessel-like field by focusing a narrow annular beam.
The electron beam propagates along the optical axis (horizontal dashed line), coaxially with the generated field.
Near the focus, the transverse radial and longitudinal components form spatially co-located $J_1^2$-type annular and $J_0^2$-type on-axis intensity profiles, respectively.
In the ideal Bessel limit, their relative intensity weights are proportional to $1-s_\theta^2$ and $s_\theta^2$.
    }
    \label{fig:radial}
\end{figure}

The radially polarized field, by contrast, contains both a transverse radial component $A_r$ and a longitudinal component $A_z$.
Because $A_r$ and $A_z$ are orthogonally polarized, no interference cross term appears in the cycle-averaged intensity, and the total intensity is given by the sum of the two component intensities.
The transverse component produces the same $J_1^2$-type annular profile as the azimuthally polarized field, whereas the longitudinal component produces a $J_0^2$-type profile with an on-axis maximum.
Figure~\ref{fig:radial} illustrates the formation of these two components by focusing a radially polarized annular beam.
In practice, a Bessel-like field is generated from an input field with a finite annular width, as implemented experimentally for tightly focused radially polarized beams \cite{Tsuru2024}.
This annulus represents the angular-spectrum distribution around the Bessel-cone angle $\theta$, whereas, in the ideal Bessel limit, its angular width approaches zero.
The relative longitudinal contribution increases with the focusing angle through the factor $s_\theta^2$ and therefore becomes significant under high-NA focusing \cite{Dorn2003,Tsuru2024}.

For later use, we introduce the propagation index $\sigma=\pm1$.
The optical phase of a Bessel component is written as
\begin{gather}
\psi^{(\sigma)} = \omega t - \sigma k\cos\theta \;z .
\label{eq:psi_sigma}
\end{gather}
Here $\sigma=+1$ denotes a co-propagating optical component, and $\sigma=-1$ denotes a counter-propagating optical component.
Along the electron trajectory,
\begin{gather}
\left(\partial_t+\beta c\partial_z\right)\psi^{(\sigma)} = \omega D^{(\sigma)} ,
\label{eq:phase_slippage}
\end{gather}
where
\begin{gather}
D^{(\sigma)} = 1-\sigma\beta\cos\theta .
\label{eq:D_sigma}
\end{gather}
This slippage factor is the optical frequency sampled by the moving electron normalized by the laboratory optical frequency, and it governs ponderomotive interactions with travelling optical waves \cite{Kozak2018NP}.

\subsubsection{Azimuthally polarized Bessel field}

We first consider the azimuthally polarized Bessel field.
In the Coulomb gauge, its vector potential is written as
\begin{gather}
    \mathbf{A}_\mathrm{az}^{(\sigma)} = A_{\mathrm{az},0}J_1(Kr)\sin\psi^{(\sigma)}\mathbf{e}_\phi .
    \label{eq:A_az}
\end{gather}
This field has no longitudinal vector-potential component. Thus the right-hand side of Eq.~(\ref{eq:G_equation_general}) vanishes, and we can choose $G=0$.
The effective ponderomotive potential is then obtained from the cycle average of the transverse vector potential:
\begin{gather}
    U_\mathrm{az}^{(\sigma)}(r) = \frac{e^2}{2\gamma m}\left\langle \left(\mathbf{A}_\mathrm{az}^{(\sigma)}\right)^2 \right\rangle .
    \label{eq:U_az_average}
\end{gather}
Using $\langle\sin^2\psi^{(\sigma)}\rangle=1/2$, we obtain
\begin{gather}
    U_\mathrm{az}^{(\sigma)}(r) = \frac{U_{\mathrm{az},0}}{\gamma}J_1^2(Kr),
    \qquad
    U_{\mathrm{az},0} = \frac{e^2A_{\mathrm{az},0}^2}{4m}.
    \label{eq:U_az_final}
\end{gather}
The result is independent of $\sigma$ because the field is purely transverse.

\subsubsection{Radially polarized Bessel field}

The radially polarized Bessel field is treated in the same framework, but it requires the gauge correction.
In the Coulomb gauge, the vector potential is written as
\begin{gather}
    \begin{aligned}
    \mathbf{A}_\mathrm{rad}^{(\sigma)}
    &=
    A_{\mathrm{rad},0}
    \left[
        \sqrt{1-s_\theta^2}J_1(Kr)\sin\psi^{(\sigma)}\mathbf{e}_r
    \right.\\
    &\left.
        -\sigma s_\theta J_0(Kr)\cos\psi^{(\sigma)}\mathbf{e}_z
    \right].
    \end{aligned}
    \label{eq:A_rad}
\end{gather}
The radial component is proportional to $\sqrt{1-s_\theta^2}$, whereas the longitudinal component,
\begin{gather}
    A_z^{(\sigma)} = -\sigma A_{\mathrm{rad},0}s_\theta J_0(Kr)\cos\psi^{(\sigma)} ,
    \label{eq:Az_rad}
\end{gather}
is proportional to $s_\theta$ and appears on the right-hand side of Eq.~(\ref{eq:G_equation_general}); the gauge function is therefore no longer zero.
Solving Eq.~(\ref{eq:G_equation_general}) gives
\begin{gather}
    G^{(\sigma)} = -\frac{\sigma \beta s_\theta}{kD^{(\sigma)}}A_{\mathrm{rad},0}J_0(Kr)\sin\psi^{(\sigma)} .
    \label{eq:G_rad}
\end{gather}
The corresponding effective longitudinal and radial components are then
\begin{gather}
    A_z^{(\sigma)}-\partial_zG^{(\sigma)} =
    -\frac{\sigma s_\theta}{D^{(\sigma)}}A_{\mathrm{rad},0}J_0(Kr)\cos\psi^{(\sigma)},
    \label{eq:Az_eff}
\end{gather}
and
\begin{gather}
    A_r^{(\sigma)}-\partial_rG^{(\sigma)} =
    \frac{\sqrt{1-s_\theta^2}-\sigma\beta}{D^{(\sigma)}}A_{\mathrm{rad},0}J_1(Kr)\sin\psi^{(\sigma)} .
    \label{eq:Ar_eff}
\end{gather}
These expressions show that the radial component is also modified by the gauge correction generated by $A_z$.

Substituting Eqs.~(\ref{eq:Az_eff}) and (\ref{eq:Ar_eff}) into Eq.~(\ref{eq:U_rel_general}), and averaging over the optical cycle, yields
\begin{gather}
    U_\mathrm{rad}^{(\sigma)}(r)
    =
    \frac{U_{\mathrm{rad},0}}{\gamma}
    \left[
        \left(1-\eta^{(\sigma)}\right)J_1^2(Kr)
        +
        \eta^{(\sigma)}J_0^2(Kr)
    \right],
    \label{eq:U_rad_final}
\end{gather}
where
\begin{gather}
    U_{\mathrm{rad},0} = \frac{e^2A_{\mathrm{rad},0}^2}{4m},
    \label{eq:U_rad0}
\end{gather}
and
\begin{gather}
    \eta^{(\sigma)}
    =
    \left( \frac{s_\theta}{\gamma D^{(\sigma)}} \right)^2.
    \label{eq:eta_rad}
\end{gather}
Thus the effective potential of the radially polarized Bessel field is a linear combination of the $J_1^2(Kr)$ and $J_0^2(Kr)$ radial profiles, with weights $1-\eta^{(\sigma)}$ and $\eta^{(\sigma)}$, respectively.

In the paraxial limit $s_\theta\to0$, one obtains $\eta^{(\sigma)}\to0$, and therefore
\begin{gather}
    U_\mathrm{rad}^{(\sigma)}(r)
    \to
    \frac{U_{\mathrm{rad},0}}{\gamma}J_1^2(Kr).
\end{gather}
In the nonrelativistic limit $\beta\to0$ and $\gamma\to1$, the weight reduces to $\eta^{(\sigma)}\to s_\theta^2$, giving
\begin{gather}
    U_\mathrm{rad}^{(\sigma)}(r)
    \to
    U_{\mathrm{rad},0}
    \left[
        \left(1-s_\theta^2\right)J_1^2(Kr)
        +
        s_\theta^2J_0^2(Kr)
    \right].
\end{gather}
In this limit, the ponderomotive potential is proportional to the local cycle-averaged optical intensity alone, and the weights reduce to the intensity fractions $1-s_\theta^2$ and $s_\theta^2$ of the transverse and longitudinal field components. In the opposite high-NA limit $s_\theta\to1$, the propagation-direction dependence disappears and $\eta^{(\sigma)}\to\gamma^{-2}$.

For the co-propagating component $\sigma=+1$, the slippage factor can become small, and $\eta^{(+)}=1$ when
\begin{gather}
    \sqrt{1-s_\theta^2}=\beta,
    \qquad
    s_\theta=\frac{1}{\gamma}.
\end{gather}
At this condition, cancellation between the radial vector potential and the gauge contribution from $A_z$ eliminates the effective radial component in Eq.~(\ref{eq:Ar_eff}), leaving a purely $J_0^2(Kr)$ Bessel potential.
This condition corresponds to slippage matching, for which the electron experiences a slowly varying optical phase and an enhanced longitudinal-field contribution.
However, the ponderomotive potential is a cycle-averaged effective potential and assumes that the electron samples many optical cycles during the interaction.
If the reduced slippage allows the electron to experience only a few cycles, this averaging is no longer quantitatively valid, and the full time-dependent interaction must be considered.

For the counter-propagating component $\sigma=-1$, the slippage factor instead increases, so no comparable enhancement occurs and the longitudinal-field contribution remains suppressed.

\subsection{Thin-lens properties for parallel-beam focusing and chromatic dispersion}

We next convert the vector-field ponderomotive potentials derived above into thin-lens properties.
Hereafter, we refer to the corresponding ponderomotive electron-optical elements as Bessel lenses.
The required scalar $J_0^2(Kr)$ and $J_1^2(Kr)$ Bessel-lens formulas are summarized in Appendix~\ref{app:bessel_lens_formulas}.
In the weak-potential thin-lens approximation, the lens power and the spherical-aberration coefficients are linear in the radial expansion coefficients of the ponderomotive potential.
Therefore, the azimuthally and radially polarized vector Bessel fields can be reduced to the corresponding scalar Bessel-lens formulas.

To keep the notation independent of the particular polarization used to generate the field, we use a common nonrelativistic ponderomotive scale $U_0$ and define
\begin{gather}
    \Lambda(\gamma)
    =
    \frac{2U_0l}{(\gamma^2-1)mc^2},
    \label{eq:Lambda}
\end{gather}
where $l$ is the effective interaction length.
This factor is obtained from the scalar formulas by using the relativistic replacement in Eq.~(\ref{eq:app_relativistic_replacement}).
In analogy with the excitation of a magnetic lens, we refer to the integrated ponderomotive strength $U_0l$ as the optical excitation of a Bessel lens.
The thin-lens approximation is appropriate for the present parameter range because the intended electron-optical focal lengths are on the millimeter scale or longer, whereas the longitudinal extent of the interaction region is typically only several tens of micrometers.

\subsubsection{Azimuthal Bessel lens}

The azimuthally polarized Bessel field has the same radial dependence as the scalar $J_1^2(Kr)$ Bessel lens.
We therefore identify its thin-lens coefficients with those of the scalar $J_1^2$ Bessel lens:
\begin{gather}
    F_\mathrm{az}
    =
    F_{J_1}
    =
    \frac{K^2}{4}\Lambda .
    \label{eq:F_az}
\end{gather}
For a parallel incident beam focused by this lens, we use $a\simeq f_\mathrm{az}$, where $f_\mathrm{az}=1/F_\mathrm{az}$.
The third- and fifth-order spherical-aberration coefficients are then
\begin{gather}
    C_{\mathrm{S3},\mathrm{az}}
    =
    -\frac{1}{2}K^2f_\mathrm{az}^3,
    \label{eq:Cs3_az_parallel}
\end{gather}
and
\begin{gather}
    C_{\mathrm{S5},\mathrm{az}}
    =
    \frac{5}{64}K^4f_\mathrm{az}^5.
    \label{eq:Cs5_az_parallel}
\end{gather}
Thus the azimuthal Bessel lens acts as a converging electron lens with negative third-order spherical aberration.

The chromatic dispersion follows directly from
$F_\mathrm{az}\propto\Lambda\propto(\gamma^2-1)^{-1}$.
This is the same inverse-squared-momentum scaling as that of the
magnetostatic solenoid lens in Eq.~(\ref{eq:FM}).
Using Eq.~(\ref{eq:abbe}), the local Abbe number is
\begin{gather}
    \nu_\mathrm{az}
    =
    -\frac{1}{2}.
    \label{eq:nu_az}
\end{gather}
The corresponding first-order chromatic-aberration coefficient is
\begin{gather}
    C_{\mathrm{c},\mathrm{az}}
    =
    \frac{2\gamma}{\gamma+1}f_\mathrm{az}.
    \label{eq:Cc_az}
\end{gather}
The azimuthal Bessel lens therefore belongs to the same dispersion class as the scalar Bessel lenses and the magnetostatic solenoid lens, for which $\nu_\mathrm{M}=-1/2$ was obtained in Eq.~(\ref{eq:abbe_M}).

\subsubsection{Radial Bessel lens}

For the radial Bessel lens, the thin-lens power is most compactly written by using the scalar $J_0^2(Kr)$ Bessel lens as the reference.
We define
\begin{gather}
    F_{J_0}
    =
    -\frac{K^2}{2}\Lambda ,
    \label{eq:F_J0}
\end{gather}
so that $F_{J_1}=-F_{J_0}/2$.
Combining this scalar relation with Eq.~(\ref{eq:U_rad_final}), the power of the radial Bessel lens is written as
\begin{gather}
    F_\mathrm{rad}^{(\sigma)}
    =
    \xi^{(\sigma)}F_{J_0},
    \label{eq:F_rad_xi}
\end{gather}
where
\begin{gather}
    \xi^{(\sigma)}
    =
    \frac{3\eta^{(\sigma)}-1}{2}.
    \label{eq:xi_eta}
\end{gather}
Here $\xi^{(\sigma)}$ is the effective strength of the radial Bessellens measured relative to the scalar $J_0^2$ Bessel lens.
Because $F_{J_0}<0$, the radial Bessel lens is diverging for $\xi^{(\sigma)}>0$ and converging for $\xi^{(\sigma)}<0$.
The zero-power condition is
\begin{gather}
    \xi^{(\sigma)}=0,
    \qquad
    \eta^{(\sigma)} = \frac{1}{3}.
    \label{eq:zero_power_condition}
\end{gather}
Thus the diverging-lens regime is reached when $\eta^{(\sigma)}>1/3$.
In the low-energy limit, this condition reduces to $s_\theta>1/\sqrt{3}$, corresponding to a high-NA regime in which the longitudinal component is sufficiently dominant.

For parallel-beam focusing, we define the signed focal length $f_\mathrm{rad}^{(\sigma)}=1/F_\mathrm{rad}^{(\sigma)}$ and use $a\simeq f_\mathrm{rad}^{(\sigma)}$.
The third-order spherical-aberration coefficient is
\begin{gather}
    \begin{aligned}
    C_{\mathrm{S3},\mathrm{rad}}^{(\sigma)}
    &=
    C_{\mathrm{S3},J_0}
    \frac{10\xi^{(\sigma)}-1}{9\xi^{(\sigma)}} \\
    &=
    -\frac{3}{8}K^2\left(f_\mathrm{rad}^{(\sigma)}\right)^3
    \frac{10\xi^{(\sigma)}-1}{9\xi^{(\sigma)}} ,
    \end{aligned}
    \label{eq:Cs3_rad_J0_ref}
\end{gather}
where $C_{\mathrm{S3},J_0}$ is the third-order spherical-aberration coefficient of a scalar $J_0^2$ Bessel lens with the same signed focal length.
The multiplicative factor accounts for the difference between the radial-vector-field expansion and the pure scalar $J_0^2$ expansion at fixed lens power.
Similarly, the fifth-order coefficient is
\begin{gather}
    \begin{aligned}
    C_{\mathrm{S5},\mathrm{rad}}^{(\sigma)}
    &=
    C_{\mathrm{S5},J_0}
    \frac{7\xi^{(\sigma)}-1}{6\xi^{(\sigma)}} \\
    &=
    \frac{5}{96}K^4\left(f_\mathrm{rad}^{(\sigma)}\right)^5
    \frac{7\xi^{(\sigma)}-1}{6\xi^{(\sigma)}} ,
    \end{aligned}
    \label{eq:Cs5_rad_J0_ref}
\end{gather}
where $C_{\mathrm{S5},J_0}$ is the fifth-order spherical-aberration coefficient of a scalar $J_0^2$ Bessel lens with the same signed focal length.

The chromatic dispersion of the radial Bessel lens consists of the scalar $J_0^2$ dispersion and the additional energy dependence of $\xi^{(\sigma)}$.
For $\xi^{(\sigma)}\neq0$, we define the logarithmic dispersion factor
\begin{gather}
    q^{(\sigma)}
    =
    \frac{d\ln \left|\xi^{(\sigma)}\right|}
    {d\ln(\gamma^2-1)} .
    \label{eq:q_rad}
\end{gather}
The absolute value is used because $\xi^{(\sigma)}$ is a signed strength, whereas $q^{(\sigma)}$ characterizes the energy scaling of its magnitude.
Since $\gamma^2-1=\beta^2\gamma^2$ is the relativistic momentum scale of the electron, $q^{(\sigma)}$ represents the logarithmic slope of the effective strength of the radial Bessel lens with respect to this scale.
The scalar reference lens has $F_{J_0}\propto(\gamma^2-1)^{-1}$, and therefore $q^{(\sigma)}$ gives the additional dispersion introduced by the longitudinal-field weight and the slippage correction.

Using Eq.~(\ref{eq:abbe}), the local Abbe number becomes
\begin{gather}
    \nu_\mathrm{rad}^{(\sigma)}
    =
    -\frac{1}{2\left(1-q^{(\sigma)}\right)} .
    \label{eq:nu_rad_q}
\end{gather}
The corresponding first-order chromatic-aberration coefficient is
\begin{gather}
    C_{\mathrm{c},\mathrm{rad}}^{(\sigma)}
    =
    \frac{2\gamma}{\gamma+1}
    \left(1-q^{(\sigma)}\right)
    f_\mathrm{rad}^{(\sigma)} .
    \label{eq:Cc_rad_q}
\end{gather}
When $q^{(\sigma)}=0$, the effective strength $\xi^{(\sigma)}$ has no additional energy scaling, and the radial Bessel lens recovers the common inverse-squared-momentum dispersion class of the scalar Bessel lenses and the magnetostatic solenoid lens,
$\nu_\mathrm{rad}^{(\sigma)}=-1/2$.

\section{Results}

\subsection{Sign tunability and self-achromatic operation of the radial Bessel lens}

We first examine the radial Bessel lens as an individual electron-optical element.
Since the lens strength scales linearly with $U_0l$, the quantities of interest are the signs of the lens power and aberration coefficients as functions of the electron energy, the cone parameter $s_\theta$, and the propagation index $\sigma$.

\begin{figure*}[t]
    \centering
    \includegraphics*[width=\textwidth]{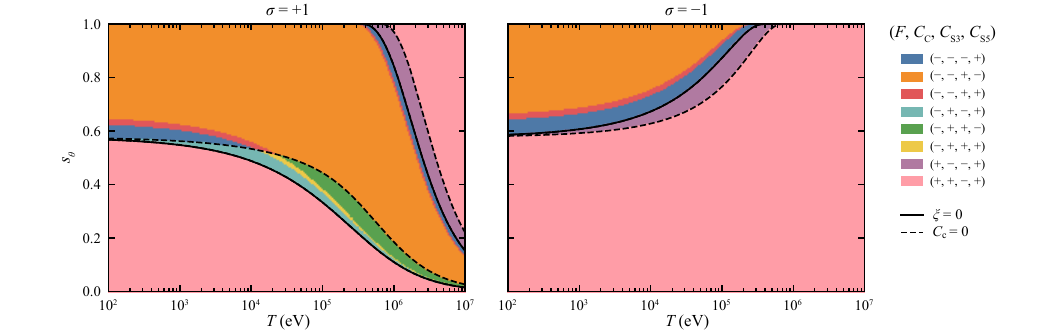}
    \caption{
    Sign map of the radial Bessel-lens properties as functions of the electron kinetic energy $T$ and the Bessel-cone parameter $s_\theta$ for $\sigma=+1$ and $\sigma=-1$.
    The colors indicate the sign combination
    $(F,C_\mathrm{c},C_{\mathrm{S3}},C_{\mathrm{S5}})$.
    The solid curve denotes the zero-power condition $\xi^{(\sigma)}=0$, and the dashed curve denotes the first-order self-achromatic condition $C_\mathrm{c}=0$, equivalently $q^{(\sigma)}=1$.
    }
    \label{fig:radial_sign_map}
\end{figure*}

Figure~\ref{fig:radial_sign_map} shows the combined sign map of
$(F,C_\mathrm{c},C_{\mathrm{S3}},C_{\mathrm{S5}})$.
The solid curve indicates the zero-power condition $\xi^{(\sigma)}=0$, while the dashed curve indicates the first-order achromatic condition $C_\mathrm{c}=0$.
The latter is equivalent to $q^{(\sigma)}=1$, for which the local Abbe number in Eq.~(\ref{eq:nu_rad_q}) diverges and the lens power becomes locally stationary at the design energy, $dF_\mathrm{rad}^{(\sigma)}/d\gamma|_{\gamma_0}=0$.
Because the $C_\mathrm{c}=0$ curve does not generally coincide with the $F=0$ curve, the radial Bessel lens can retain a finite lens power while exhibiting zero first-order chromatic aberration.
This property originates from the additional energy dependence of
$\xi^{(\sigma)}$, which is absent from the scalar Bessel and magnetostatic dispersion class with $\nu=-1/2$.
The radial Bessel lens can therefore operate as a first-order self-achromatic single lens rather than requiring a second element with a different dispersion.

The map also shows that the signs of the geometrical aberrations are not fixed by the sign of the lens power: the third- and fifth-order spherical aberrations change sign at $\xi^{(\sigma)}=1/10$ and $\xi^{(\sigma)}=1/7$, respectively, conditions distinct from both the zero-power and zero-chromatic-aberration conditions.
Consequently, converging and diverging radial Bessel lenses can each occur with multiple combinations of
$C_\mathrm{c}$, $C_{\mathrm{S3}}$, and $C_{\mathrm{S5}}$.

The two propagation branches differ markedly.
For $\sigma=+1$, all eight sign combinations of $(F,C_\mathrm{c},C_{\mathrm{S3}},C_{\mathrm{S5}})$ appear in Fig.~\ref{fig:radial_sign_map}.
The $C_\mathrm{c}=0$ condition follows the low-$s_\theta$ region at low energies, and a second $C_\mathrm{c}=0$ branch emerges as the energy increases, the two branches together spanning a wide range of small and large cone parameters.
This richness reflects the slippage-matched enhancement of the longitudinal response on the co-propagating branch, and makes $\sigma=+1$ the more varied regime of the relativistic light--electron interaction.
For $\sigma=-1$, five of the eight combinations occur, and the single $C_\mathrm{c}=0$ curve passes through a positive-power region with negative third-order and positive fifth-order spherical aberration: a single radial Bessel lens can there simultaneously provide converging power, zero first-order chromatic aberration, and negative $C_{\mathrm{S3}}$.
 
Scalar ponderomotive lenses are already distinctive among electron lenses in offering negative power or negative third-order spherical aberration.
The radial vector field goes further: of these eight sign combinations, only $(-,-,+,-)$ and $(+,+,-,+)$ are available to scalar Bessel lenses.
The remaining six, including converging lenses with zero or negative $C_\mathrm{c}$ and diverging lenses with negative $C_{\mathrm{S3}}$, arise only from the relativistically corrected vector field.

\subsection{Simultaneous correction of chromatic and third-order spherical aberrations}

We next examine zero-separation doublets that simultaneously cancel first-order chromatic aberration and third-order spherical aberration while retaining a prescribed objective-lens power.
The doublet combines a reference lens of power $F_\mathrm{ref}$, which is either the magnetic objective-lens model or an azimuthal Bessel lens, with a radial Bessel lens of power $F_\mathrm{rad}$.
The total power is fixed, $F_\mathrm{tot}=F_\mathrm{ref}+F_\mathrm{rad}=1/f_\mathrm{tot}$, so that the choice of $F_\mathrm{ref}$ determines $F_\mathrm{rad}=F_\mathrm{tot}-F_\mathrm{ref}$.
The optical excitation of the radial Bessel lens required for this power, denoted $(U_0l)_\mathrm{rad}$, follows from Eqs.~(\ref{eq:Lambda}), (\ref{eq:F_J0}), and (\ref{eq:F_rad_xi}):
\begin{equation}
    (U_0l)_\mathrm{rad}
    =
    -
    \frac{
    (\gamma^2-1)mc^2
    \left(
    F_\mathrm{tot}-F_\mathrm{ref}
    \right)
    }{
    K_\mathrm{rad}^2
    \xi^{(\sigma)}
    },
    \label{eq:radial_excitation_fixed_power}
\end{equation}
where $K_\mathrm{rad}=(2\pi/\lambda_\mathrm{rad})s_{\theta,\mathrm{rad}}$ is the transverse wavenumber of the radial Bessel field, with $\lambda_\mathrm{rad}$ and $s_{\theta,\mathrm{rad}}$ its optical wavelength and cone parameter; $K_\mathrm{az}$, $\lambda_\mathrm{az}$, and $s_{\theta,\mathrm{az}}$ are defined analogously for the azimuthal Bessel lens.

\begin{figure*}[t]
    \centering
    \includegraphics[width=\textwidth]{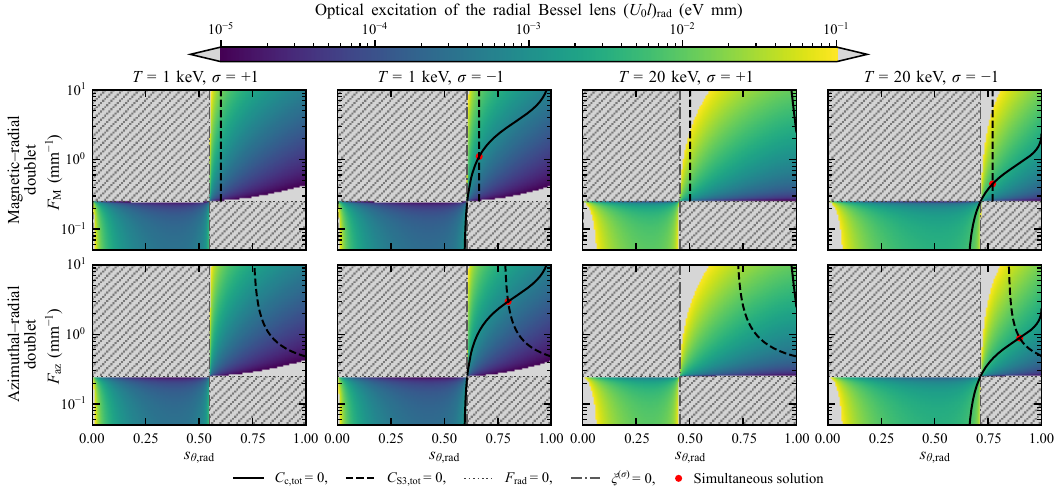}
    \caption{
    Design maps for simultaneous correction of first-order chromatic aberration and third-order spherical aberration at fixed total power ($f_\mathrm{tot}=4$ mm).
    The radial Bessel lens is combined with the magnetic objective-lens model (upper row) or an azimuthal Bessel lens (lower row); the vertical axis is the reference-lens power, and the horizontal axis is the radial cone parameter.
    The optical wavelengths of the radial and azimuthal Bessel fields are both $1064~\mathrm{nm}$, and $s_{\theta,\mathrm{az}}=0.6$.
    The color scale gives the required radial excitation $(U_0l)_\mathrm{rad}$; hatched regions would require $(U_0l)_\mathrm{rad}<0$, and gray regions lie outside the displayed excitation range.
    Line styles and the simultaneous-solution markers are identified in the legend.
    }
    \label{fig:result_2}
\end{figure*}

The first-order chromatic condition follows from Eq.~(\ref{eq:ach}),
\begin{equation}
    \frac{F_\mathrm{ref}}{\nu_\mathrm{ref}}
    +
    \frac{F_\mathrm{rad}}{\nu_\mathrm{rad}}
    =
    0.
    \label{eq:doublet_achromat_condition}
\end{equation}
For the spherical condition, we describe each thin lens by the transverse ``kick'' it imparts to a ray at height $r$, that is, the change in trajectory slope, or equivalently the transverse-momentum transfer normalized by the axial momentum:
\begin{equation}
    \Theta(r)
    =
    -Fr+\Theta_3r^3+O(r^5),
    \qquad
    \Theta_3=-C_\mathrm{S3}F^4,
    \label{eq:thin_lens_kick}
\end{equation}
where the expression for $\Theta_3$ follows from the definition of the spherical-aberration coefficient.
For a zero-separation doublet the kicks add; because $F_\mathrm{tot}$ is fixed and nonzero, $C_\mathrm{S3,tot}=0$ is equivalent to
\begin{equation}
    \Theta_{3,\mathrm{ref}}
    +
    \Theta_{3,\mathrm{rad}}
    =
    0.
    \label{eq:doublet_spherical_condition}
\end{equation}

The upper row of Fig.~\ref{fig:result_2} uses the magnetic objective-lens model as the reference: its power $F_\mathrm{M}$ is varied, its dispersion is $\nu_\mathrm{M}=-1/2$, and its third-order coefficient follows $C_\mathrm{S3,M}/f_\mathrm{M}=1$, a ratio representative of well-designed conventional objective lenses \cite{JEOLSpherical}, giving $\Theta_{3,\mathrm{M}}=-F_\mathrm{M}^3$.
The lower row uses an azimuthal Bessel lens as the reference.
Its local Abbe number equals that of the magnetic objective-lens model, $\nu_\mathrm{az}=\nu_\mathrm{M}=-1/2$, so the chromatic condition is identical in the two rows; they differ only through the third-order kick coefficient, which for the azimuthal Bessel lens follows from Eq.~(\ref{eq:Cs3_az_parallel}) as $\Theta_{3,\mathrm{az}}=(K_\mathrm{az}^2/2)F_\mathrm{az}$.

Figure~\ref{fig:result_2} shows the resulting design maps at 1 and 20 keV for both propagation indices, calculated for $f_\mathrm{tot}=4$~mm, $\lambda_\mathrm{rad}=\lambda_\mathrm{az}=1064$~nm, and $s_{\theta,\mathrm{az}}=0.6$.
The color scale gives the required radial excitation $(U_0l)_\mathrm{rad}$ of Eq.~(\ref{eq:radial_excitation_fixed_power}), and the solid and dashed contours indicate $C_\mathrm{c,tot}=0$ and $C_\mathrm{S3,tot}=0$; their intersections give the simultaneous solutions.
The horizontal dotted line marks $F_\mathrm{rad}=0$, and the vertical dash-dotted line marks $\xi^{(\sigma)}=0$, where the power of the radial Bessel lens per unit excitation vanishes and changes sign.
Since the optical excitation must satisfy $(U_0l)_\mathrm{rad}>0$, the physically accessible regions obey $\xi^{(\sigma)}(F_\mathrm{tot}-F_\mathrm{ref})<0$: the two boundary lines divide each panel into four domains, of which only two diagonally opposite domains are physical.
The complementary hatched regions would require negative optical excitation, and the required excitation diverges on approaching $\xi^{(\sigma)}=0$ at finite $F_\mathrm{rad}$.

For the parameters shown, physical simultaneous solutions are obtained only for $\sigma=-1$; for the azimuthal--radial doublet, this statement refers to the fixed azimuthal cone parameter $s_{\theta,\mathrm{az}}=0.6$.
Appendix~\ref{app:saz_scan} extends these maps to other azimuthal cone parameters and to 200 keV: reducing $s_{\theta,\mathrm{az}}$ moves the physical solution to smaller $s_{\theta,\mathrm{rad}}$ and weaker component powers, while a higher electron energy pushes it toward $s_{\theta,\mathrm{rad}}\to1$.
At 200 keV, a physical solution also appears for $\sigma=+1$, although only at extreme radial cone parameters, $s_{\theta,\mathrm{rad}}\gtrsim0.98$.

The corresponding simultaneous-solution parameters are summarized in Table~\ref{tab:fixed_total_doublet_solutions}.
The magnetic--radial doublet requires less cancellation between the individual lens powers than the azimuthal--radial doublet.
This difference is particularly pronounced at 1 keV, where the all-Bessel design combines
$F_\mathrm{az}=2.97~\mathrm{mm}^{-1}$
and
$F_\mathrm{rad}=-2.72~\mathrm{mm}^{-1}$
to produce the much smaller net power
$F_\mathrm{tot}=0.25~\mathrm{mm}^{-1}$.
Although both
$C_\mathrm{c,tot}$ and
$C_\mathrm{S3,tot}$ vanish at the marked intersections, the strong cancellation between the individual lens powers increases the importance of residual higher-order aberrations.

\begin{table}[t]
    \centering
    \caption{
    Simultaneous solutions of
    $f_\mathrm{tot}=4$ mm,
    $C_\mathrm{c,tot}=0$, and
    $C_\mathrm{S3,tot}=0$
    in Fig.~\ref{fig:result_2}.
    The optical wavelengths of the radial and azimuthal Bessel fields are both $1064~\mathrm{nm}$.
    The radial Bessel lens uses $\sigma=-1$, and the azimuthal Bessel lens uses
    $s_{\theta,\mathrm{az}}=0.6$.
    }
    \label{tab:fixed_total_doublet_solutions}
    \begingroup
    \setlength{\tabcolsep}{1.8pt}
    \begin{ruledtabular}
        \begin{tabular}{lcccc}
            & \multicolumn{2}{c}{Magnetic--radial}
            & \multicolumn{2}{c}{Azimuthal--radial}\\
            $T$ (keV) & 1 & 20 & 1 & 20\\
            \hline
            $s_{\theta,\mathrm{rad}}$
            & 0.663 & 0.771 & 0.798 & 0.898\\
            \multicolumn{5}{c}{Lens power $F$ (mm$^{-1}$)}\\
            Reference
            & 1.11 & 0.448 & 2.97 & 0.887\\
            Radial
            & $-0.861$ & $-0.198$ & $-2.72$ & $-0.637$\\
            \hline
            \multicolumn{5}{c}{Optical excitation $U_0l$ (eV mm)}\\
            Reference
            & --- & --- & $9.46\times10^{-4}$ & $5.76\times10^{-3}$\\
            Radial
            & $1.12\times10^{-3}$ & $3.89\times10^{-3}$ & $6.36\times10^{-4}$ & $2.35\times10^{-3}$\\
        \end{tabular}
    \end{ruledtabular}
    \endgroup
\end{table}

\subsection{Transverse-aberration comparison with conventional objective lenses}

\begin{figure*}[t]
    \centering
    \includegraphics[width=\textwidth]{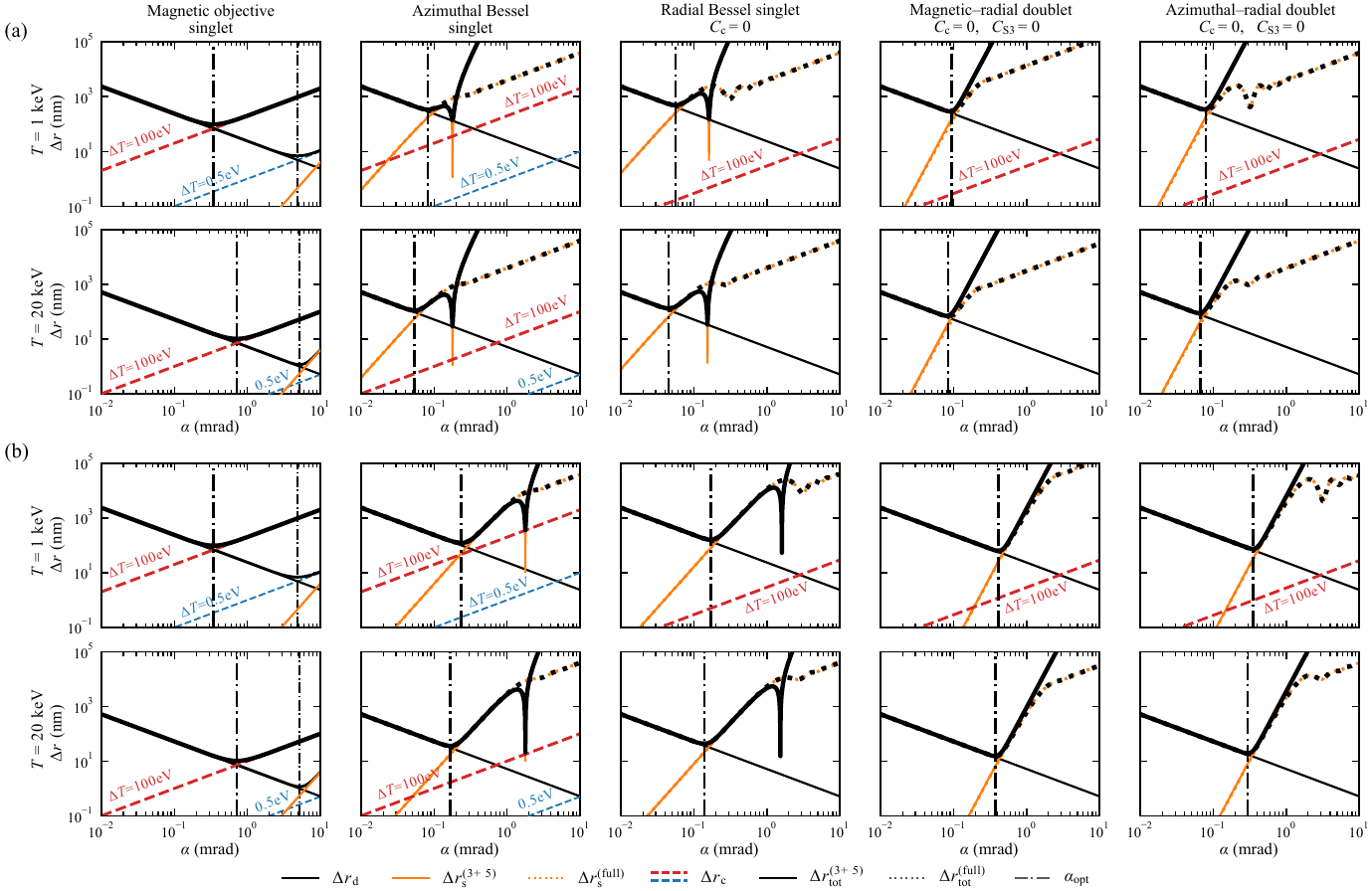}
    \caption{
    Transverse-aberration comparison of the five configurations of Table~\ref{tab:aberration_optima}, with optical wavelengths of (a) $1064~\mathrm{nm}$ and (b) $10.6~\mathrm{\mu m}$ for both Bessel fields.
    Model parameters and correction conditions are given in the text.
    Panels containing a Bessel lens show both the third- plus fifth-order and full-Bessel spherical and total radii; curve styles, energy widths, and the optimum semi-angles are identified in the legend.
    }
    \label{fig:result_3}
\end{figure*}

Figure~\ref{fig:result_3} compares the finite-aperture performance of five configurations with a common $f_\mathrm{tot}=4$ mm: a magnetic objective-lens model, an azimuthal Bessel singlet, a self-achromatic radial Bessel singlet, and the magnetic--radial and azimuthal--radial doublets satisfying $C_\mathrm{c,tot}=C_\mathrm{S3,tot}=0$.
The two rows correspond to electron energies of 1 and 20 keV, and each panel includes energy widths of
$\Delta T=0.5$ and 100 eV.
The definitions of the diffraction, spherical, chromatic, and total transverse radii are given in Appendix~\ref{app:transverse_aberration_evaluation}.
The azimuthal singlet uses $s_{\theta,\mathrm{az}}=0.6$.
The self-achromatic radial singlet uses $\sigma=-1$ and $q^{(-)}=1$, giving $s_{\theta,\mathrm{rad}}=0.592$ at 1 keV and 0.652 at 20 keV.

For the magnetic objective-lens model, we use
\begin{equation}
    \frac{C_\mathrm{S3,M}}{f_\mathrm{M}}
    =
    1,
    \qquad
    \frac{C_\mathrm{S5,M}}{f_\mathrm{M}}
    =
    5,
    \qquad
    \nu_\mathrm{M}
    =
    -\frac12,
    \label{eq:magnetic_aberration_model_result}
\end{equation}
which gives $C_\mathrm{S3,M}=4$ mm and $C_\mathrm{S5,M}=20$ mm at $f_\mathrm{M}=f_\mathrm{tot}$, together with the chromatic coefficient calculated consistently from the local Abbe number,
\begin{equation}
    C_\mathrm{c,M}
    =
    -\frac{\gamma}
    {(\gamma+1)F_\mathrm{M}\nu_\mathrm{M}}
    =
    \frac{2\gamma}{\gamma+1}
    f_\mathrm{M}.
    \label{eq:magnetic_Cc_result}
\end{equation}
The resulting $C_\mathrm{c,M}\approx f_\mathrm{M}$ is close to typical objective-lens values; practical coefficients vary with the field profile, polepiece geometry, and imaging conjugates and need not follow this dispersion-derived relation.

For the radial Bessel singlet and both corrected doublets, the first derivative of the total power vanishes at the design energy.
Their finite-bandwidth chromatic radius is therefore evaluated from the complete energy dependence of the lens power:
\begin{equation}
    \Delta r_\mathrm{c}
    =
    \alpha f_\mathrm{tot}
    \max_{T_-\leq T\leq T_+}
    \left|
    1
    -
    \frac{F_\mathrm{tot}(T)}
    {F_\mathrm{tot}(T_0)}
    \right|.
    \label{eq:residual_chromatic_radius}
\end{equation}
This retains the second- and higher-order energy dependence that remains after the first-order condition
$C_\mathrm{c}=0$ has been imposed.

For the doublets, the spherical-aberration contributions are combined through the kick coefficients of Sec.~III\,B, extended to fifth order as $\Theta_5=-C_\mathrm{S5}F^6$:
\begin{equation}
    \Theta_{3,\mathrm{tot}}
    =
    \sum_i \Theta_{3,i},
    \qquad
    \Theta_{5,\mathrm{tot}}
    =
    \sum_i \Theta_{5,i},
    \label{eq:doublet_spherical_combination_result}
\end{equation}
The composite coefficients are then
\begin{equation}
    C_\mathrm{S3,tot}
    =
    -
    \frac{\Theta_{3,\mathrm{tot}}}
    {F_\mathrm{tot}^4},
    \qquad
    C_\mathrm{S5,tot}
    =
    -
    \frac{\Theta_{5,\mathrm{tot}}}
    {F_\mathrm{tot}^6}.
    \label{eq:doublet_composite_Cs_result}
\end{equation}

In addition to the third- plus fifth-order approximation, all panels containing a Bessel lens show a non-expanded evaluation of the retained ideal Bessel profile.
The model assumes that the Bessel field is independent of $z$ over an effective length $l$ and is abruptly truncated outside this interval.
The longitudinal integration then changes only the excitation scale:
\begin{equation}
    \int U(r,z)dz
    =
    lU(r).
    \label{eq:ideal_Bessel_projection}
\end{equation}
For an azimuthal Bessel element, the non-expanded thin-lens kick is
\begin{equation}
    \Theta_\mathrm{az}^{(\mathrm{full})}(r)
    =
    -
    \frac{2F_\mathrm{az}}{K_\mathrm{az}}
    \left.
    \frac{dJ_1^2(x)}{dx}
    \right|_{x=K_\mathrm{az}r}.
    \label{eq:full_azimuthal_kick_result}
\end{equation}
For a radial Bessel element, it is
\begin{multline}
    \Theta_\mathrm{rad}^{(\mathrm{full})}(r)
    =
    \frac{F_\mathrm{rad}^{(\sigma)}}
    {\xi^{(\sigma)}K_\mathrm{rad}}
    \\
    \times
    \left.
    \frac{d}{dx}
    \left[
    \left(
    1-\eta^{(\sigma)}
    \right)
    J_1^2(x)
    +
    \eta^{(\sigma)}
    J_0^2(x)
    \right]
    \right|_{x=K_\mathrm{rad}r}.
    \label{eq:full_radial_kick_result}
\end{multline}
Both expressions are normalized so that their linear terms equal
$-F r$.
The non-expanded spherical displacement at the nominal focal plane is obtained directly from the total kick:
\begin{equation}
    \Delta r_\mathrm{s}^{(\mathrm{full})}
    =
    \left|
    -f_\mathrm{tot}
    \left[
    \Theta_\mathrm{tot}
    \left(
    f_\mathrm{tot}\alpha
    \right)
    +
    F_\mathrm{tot}
    f_\mathrm{tot}\alpha
    \right]
    \right|.
    \label{eq:full_Bessel_spherical_radius}
\end{equation}
In the magnetic--radial doublet, the magnetic contribution retains its third- and fifth-order terms.

Figure~\ref{fig:result_3}(a) uses an optical wavelength of $1064~\mathrm{nm}$ for both Bessel fields.
The optimum semi-angles marked in the panels follow the cumulative pupil criterion of Appendix~\ref{app:transverse_aberration_evaluation}, applied to the full-Bessel evaluation where available and to the third- plus fifth-order model for the magnetic singlet.
Table~\ref{tab:aberration_optima} summarizes the optimum semi-angles, total radii, and optical excitations for all configurations at both optical wavelengths.

\begin{table*}[t]
    \caption{
    Optimum semi-angle $\alpha_\mathrm{opt}$ (mrad), total transverse radius $\Delta r_\mathrm{tot}$ (nm), and optical excitation $U_0l$ (eV mm) for the configurations of Fig.~\ref{fig:result_3} ($f_\mathrm{tot}=4$ mm).
    Bessel-containing configurations use the full-Bessel evaluation at $\Delta T=100$ eV; their optima are nearly independent of $\Delta T$ (the largest change at $\Delta T=0.5$ eV is the azimuthal singlet at $10.6~\mathrm{\mu m}$ and 1 keV, 113 nm).
    For the azimuthal--radial doublet, the ordered pair in the $U_0l$ column gives $((U_0l)_\mathrm{az},(U_0l)_\mathrm{rad})$; all other entries refer to the single optical element.
    The magnetic objective-lens model is independent of optical wavelength and is listed for both energy widths.
    }
    \label{tab:aberration_optima}
    \begin{ruledtabular}
        \begin{tabular}{lcccccc}
        & \multicolumn{3}{c}{$T=1$ keV} & \multicolumn{3}{c}{$T=20$ keV}\\
        Configuration
        & $\alpha_\mathrm{opt}$
        & $\Delta r_\mathrm{tot}$
        & $U_0l$
        & $\alpha_\mathrm{opt}$
        & $\Delta r_\mathrm{tot}$
        & $U_0l$\\
        \hline
        Magnetic objective, $\Delta T=0.5$ eV & 4.83 & 6.90 & --- & 5.20 & 1.18 & ---\\
        Magnetic objective, $\Delta T=100$ eV & 0.344 & 97.3 & --- & 0.717 & 10.3 & ---\\
        \hline
        \multicolumn{7}{l}{$\lambda_\mathrm{rad}=\lambda_\mathrm{az}=1.064~\mu$m:}\\
        Azimuthal singlet & 0.0820 & $3.40\times10^2$ & $7.97\times10^{-5}$ & 0.0540 & 113 & $1.62\times10^{-3}$\\
        Radial singlet ($C_\mathrm{c}=0$) & 0.0558 & 495 & $1.67\times10^{-3}$ & 0.0452 & 135 & $7.30\times10^{-3}$\\
        Magnetic--radial doublet & 0.0934 & 280 & $1.12\times10^{-3}$ & 0.0844 & 68.7 & $3.89\times10^{-3}$\\
        Azimuthal--radial doublet & 0.0792 & 330 & $(9.46,6.36)\times10^{-4}$ & 0.0666 & 86.9 & $(5.76,2.35)\times10^{-3}$\\
        \hline
        \multicolumn{7}{l}{$\lambda_\mathrm{rad}=\lambda_\mathrm{az}=10.6~\mu$m:}\\
        Azimuthal singlet & 0.235 & 122 & $7.91\times10^{-3}$ & 0.166 & 36.5 & $1.61\times10^{-1}$\\
        Radial singlet ($C_\mathrm{c}=0$) & 0.170 & 161 & $1.65\times10^{-1}$ & 0.139 & 43.6 & $7.25\times10^{-1}$\\
        Magnetic--radial doublet & 0.415 & 62.5 & $1.12\times10^{-1}$ & 0.373 & 15.4 & $3.86\times10^{-1}$\\
        Azimuthal--radial doublet & 0.350 & 74.1 & $(9.39,6.31)\times10^{-2}$ & 0.298 & 19.3 & $(5.72,2.33)\times10^{-1}$\\
        \end{tabular}
    \end{ruledtabular}
\end{table*}

At $1064~\mathrm{nm}$, substantial finite-aperture spherical aberration remains even though the corrected doublets satisfy
$C_\mathrm{c,tot}=C_\mathrm{S3,tot}=0$: the short transverse scale of the Bessel field produces dominant fifth- and higher-order nonlinearities over the usable pupil, and the optima of all Bessel-containing configurations are set by diffraction and residual spherical aberration, nearly independently of $\Delta T$.
The effectiveness of the chromatic correction itself is nevertheless clear.
For the self-achromatic radial singlet and both corrected doublets, the $\Delta T=0.5$ eV chromatic curves at 1 keV and the $\Delta T=100$ eV curves at 20 keV lie below the plotted range.
At $1064~\mathrm{nm}$, the corrected doublets remain far above the chromatically limited magnetic objective-lens model for either energy width because of their residual geometrical aberration.

Figure~\ref{fig:result_3}(b) repeats the calculation with the optical wavelength of both Bessel fields increased to
10.6~$\mu$m.
All Bessel lenses are redesigned while retaining the same electron-optical focal length and correction conditions.
The longer optical wavelength broadens the transverse Bessel profile, so the electron pupil samples a more nearly linear region of the kick; the optimum radii of all Bessel-containing configurations decrease by factors of approximately three to five (Table~\ref{tab:aberration_optima}).
For $\Delta T=100$ eV at 1 keV, both corrected doublets now outperform the magnetic objective-lens model.
For $\Delta T=0.5$ eV, however, the magnetic objective-lens model remains substantially better because its chromatic broadening is already small, and at 20 keV it remains superior for both energy widths.
The corrected Bessel doublets are therefore most favorable in the low-energy, large-energy-spread regime.

\section{Discussion}

The constrained doublet search shows that a radial Bessel lens, combined with either a magnetic or an azimuthal reference lens, provides enough control over dispersion and geometrical aberration to cancel the first-order chromatic and third-order spherical-aberration coefficients simultaneously, at a fixed total focal length of $f_\mathrm{tot}=4$~mm.
Three questions follow from this result: whether the required optical fields are experimentally accessible, where such a corrector would be useful beyond monochromatized imaging, and what the present analysis leaves unresolved.

\subsection{Optical requirements and practical implementation}

\subsubsection{Wavelength scaling of the required optical power and lens properties}

The integrated ponderomotive strength $U_0l$ links the electron-optical design to the required laser power.
For the narrow-annulus model of Appendix~\ref{app:bessel_power_estimate}, the total power carried by the Bessel-like field is
\begin{equation}
    P_\mathrm{tot}
    =
    \frac{\lambda U_0l}{2\pi\mu},
    \qquad
    \mu
    =
    \frac{h\alpha_\mathrm{fs}}{m\omega^2}.
    \label{eq:discussion_power_from_Q}
\end{equation}
In the $\mathrm{eV\,mm}$ unit adopted in the Results, the two wavelengths of Fig.~\ref{fig:result_3} give
\begin{align}
    \frac{P_\mathrm{tot}}{U_0l}
    &\simeq
    16.0~\mathrm{MW\,(eV\,mm)^{-1}}
    &&(\lambda=1064~\mathrm{nm}),
    \\
    \frac{P_\mathrm{tot}}{U_0l}
    &\simeq
    1.61~\mathrm{MW\,(eV\,mm)^{-1}}
    &&(\lambda=10.6~\mathrm{\mu m}),
    \label{eq:discussion_power_reference}
\end{align}
so the optical excitations in Table~\ref{tab:aberration_optima} convert directly into optical powers.
At $T=1~\mathrm{keV}$ and $\lambda=1064~\mathrm{nm}$, the magnetic--radial doublet requires $17.9~\mathrm{kW}$ in the radial Bessel field, whereas the azimuthal--radial doublet requires $15.1$ and $10.2~\mathrm{kW}$ in the azimuthal and radial fields, respectively.
At $\lambda=10.6~\mathrm{\mu m}$ the corresponding values are approximately $180$, $151$, and $102~\mathrm{kW}$.

This difference follows from the wavelength dependence at fixed cone parameter and electron-lens power,
\begin{equation}
    U_0l\propto\lambda^2,
    \qquad
    P_\mathrm{tot}\propto\lambda,
    \qquad
    \Theta_3\propto\lambda^{-2},
    \qquad
    \Theta_5\propto\lambda^{-4}.
    \label{eq:Bessel_wavelength_scaling_result}
\end{equation}
A longer wavelength suppresses the finite-aperture nonlinear kick at the cost of a higher optical power.
Moving from $1064~\mathrm{nm}$ to $10.6~\mathrm{\mu m}$ raises the required optical power by approximately one order of magnitude while reducing the third- and fifth-order nonlinear kick coefficients by approximately two and four orders of magnitude, respectively.
This suppression appears directly in the design performance.
As shown in Fig.~\ref{fig:result_3}(b), both corrected doublets outperform the magnetic objective-lens model at $10.6~\mathrm{\mu m}$ for $T=1~\mathrm{keV}$ and $\Delta T=100~\mathrm{eV}$.

The wavelength cannot be increased indefinitely.
The cycle-averaged ponderomotive description requires the optical oscillation sampled along the electron trajectory, whose angular frequency scales as $\omega D^{(\sigma)}$, to remain fast relative to the field-envelope and electron-transit timescales \cite{Kibble1966,Startsev1997}.
The wavelength is therefore set by a balance among optical power, residual geometrical aberration, available laser technology, and the validity of the cycle-averaged model.

\subsubsection{Configuration of the all-optical doublet}

The zero-separation azimuthal--radial doublet can be built with a single optical focusing system.
Two concentric annular optical beams, one radially and one azimuthally polarized, are sent into the entrance pupil of a common optical objective lens.
Their annular radii independently set $s_{\theta,\mathrm{rad}}$ and $s_{\theta,\mathrm{az}}$, while their optical powers set the corresponding values of $U_0l$.
The objective lens then produces spatially co-located radial and azimuthal Bessel-like fields on the electron axis.
For the useful counter-propagating solutions, both beams are launched in the direction opposite to the electron propagation.

The geometry requires a large NA together with an electron-beam through-hole on the optical axis.
In the visible and near-infrared, a reflective objective lens with an NA of $0.6$ and a through-hole of about $1~\mathrm{mm}$ in diameter in its central secondary mirror is feasible.
Because the entrance pupil is filled by a narrow annulus, an aperture of this size in the secondary mirror does not vignette the annular beams.
An objective lens with a higher NA would cover the low and intermediate $s_\theta$ regions of the present design maps through the choice of annular radius, giving access to the useful operating points.

\subsubsection{Feasibility of the optical implementation}

For pulsed operation, the peak power required near $1~\mathrm{\mu m}$ is already below demonstrated values.
Mihaila \textit{et al.} used $1035$-nm, $280$-fs pulses with energies of $2.5$ and $3.1~\mu\mathrm{J}$ for concave and convex ponderomotive electron lenses, respectively, reaching focal lengths down to $1.6~\mathrm{mm}$ \cite{Mihaila2022}.
The peak powers estimated from the pulse energy and duration are $8.9$ and $11.1~\mathrm{MW}$.
Their focused-pulse geometry does not translate directly into the present $U_0l$, but these results show that the tens-of-kW scale required by the $1064$-nm designs is compatible with an ultrashort-pulse experiment synchronized to the electron pulse.

The $10.6~\mathrm{\mu m}$ band is attractive because high-power carbon-dioxide (CO$_2$) lasers are available.
Their average power is sufficient in comparison with that of $1~\mathrm{\mu m}$ sources, but no mechanism equivalent to a mode-locked short-pulse oscillator has been established at this wavelength, so comparable peak powers are not yet accessible.
Nanosecond- or microsecond-scale pulses may reach the required peak power, but they typically show an irregular, spiked temporal envelope rather than a smooth one.
Controlling this envelope is necessary before such a pulse can act as an effective electron lens, and this remains an open issue.

An alternative to the use of pulsed peak power is continuous-wave operation combined with an enhancement cavity.
At $1~\mathrm{\mu m}$ this approach is well established.
A near-concentric cavity produced $7.5$~kW at a $7$-$\mu$m waist \cite{Schwartz2017}, an electron phase-plate system was designed for approximately $97$~kW of one-way circulating power with a buildup exceeding $7\times10^3$ \cite{Turnbaugh2021}, and a high-finesse cavity sustained up to $710$~kW of average intracavity power \cite{Lu2024}.
For CO$_2$ lasers the demonstrated buildup factors remain modest; Ando \textit{et al.} obtained an approximately $50$-fold buildup at $10.23~\mathrm{\mu m}$ \cite{Ando2013}.

Enhancement cavities nevertheless present substantial difficulties.
First, high-NA focusing inside a cavity is technically demanding.
Second, although Bessel-Gaussian cavity concepts have been proposed \cite{Putnam2012}, no cavity sustaining such a mode at high power has been demonstrated.
In addition, a round-trip cavity forms an optical standing wave, so both the $\sigma=+1$ and $\sigma=-1$ branches contribute and must be treated together.
The $\sigma=+1$ branch is the more critical of the two because $D^{(+)}$ can become small, approaching the slippage-matched regime in which the cycle averaging itself breaks down.
A design aimed at achromatic correction therefore requires a more detailed treatment of this point.

\subsection{Achromatic optics for energy-preserving imaging}

Monochromatization reduces the effect of chromatic aberration, but is not available when a finite energy distribution is an essential part of the electron-beam state.
Radio-frequency acceleration generally produces an energy--time correlation, and attosecond bunching schemes require an energy modulation that is subsequently converted into temporal structure.
In the inelastic ponderomotive-scattering experiment of Koz\'ak \textit{et al.}, $29$-keV electrons acquired an energy spread of $19.6$~keV, so a broad energy distribution was generated intentionally \cite{Kozak2018NP}.
Such beams must be focused for excitation or imaging while their broad longitudinal phase-space distribution is preserved, and chromatic-aberration correction is required for this purpose.

Mihaila \textit{et al.} proposed a time-dependent ponderomotive lens that compensates the correlation between electron energy and arrival time \cite{Mihaila2025PRL}.
That dynamically synchronized method differs from the static dispersion engineering considered here.
Both approaches nevertheless rely on degrees of freedom specific to optical fields, namely a high temporal resolution and a field distribution that can be designed through polarization and wavefront.
Correction that leaves a functional energy distribution intact becomes possible only through these degrees of freedom.

Electron-beam components of different energy may also serve as measurement variables.
In optical microscopy, stimulated-emission-depletion and multicolor fluorescence methods use different optical frequencies for distinct functions or specimen constituents \cite{Hell1994,Chen2021}.
By analogy, an electron-optical system might transport two or more energy components to a common specimen or image plane while keeping them distinguishable in electron-beam energy.
Such a system would support energy-selective excitation, imaging, or detection within a single column.
An achromatic imaging column may therefore open the way to electron-optical measurements that have not yet been realized.

\subsection{Extended design freedom and remaining issues}

The classification of dispersion classes and the design maps established here provide a basis for extending the design freedom and searching for the best trade-off against higher-order spherical aberration.
All of the extensions discussed below can be treated within the framework developed in this work.

One additional degree of freedom is dispersive.
Combining the radial ponderomotive lens with an electrostatic lens, which belongs to a different dispersion class, enlarges the set of available correction conditions.
Another is geometrical.
The present study treated only zero-separation doublets, whereas separating the component lenses longitudinally, or using an electron-optical relay to control the ray height, would redistribute the electron-lens power and the geometrical aberration among the elements.

The choice of optical-field geometry also affects the design and implementation freedom.
Two crossed cylindrical ponderomotive lenses can produce an approximately round correction \cite{Uesugi2025}.
An optical beam crossing the electron axis can produce an electric-field component parallel to the electron velocity without relying on the longitudinal field generated by high-NA focusing.
Such a configuration would relax the optical constraints of the coaxial Bessel geometry while retaining the longitudinal-field response on which the correction depends.
The higher-order transverse kick that limits the present designs is intrinsic to the Bessel profile.
Reducing the residual spherical aberration without recourse to a longer optical wavelength therefore requires a field that combines a strong and controllable longitudinal component with a more linear transverse kick.
Identifying such vector fields remains an open problem in optical design.

Several effects are not included in the present model and must be evaluated alongside these extensions.
First, the third-order spherical-aberration coefficient is canceled only at the design energy.
Away from this energy the component kicks no longer cancel exactly, as the finite-energy curves in Fig.~\ref{fig:result_3} show, and this dependence must be quantified when the accepted energy interval is large.
Second, an optical arrangement that forms a standing wave introduces a simultaneous contribution from the $\sigma=+1$ branch, and the validity of the cycle averaging must be examined in that regime.
Third, off-axis aberrations such as coma must be included in the design once the corrector is incorporated into a complete imaging column.

\section{Conclusion}

We have developed a relativistic electron-optical description of vector Bessel ponderomotive lenses.
A radially polarized Bessel field contains a transverse optical-field component and an on-axis component polarized parallel to the electron velocity.
Because these components have different relativistic responses, their relative electron-lens contributions depend on the electron energy, the Bessel cone angle, and whether the light and the electrons propagate in the same or in opposite directions.
This dependence gives control over both the sign of the electron-lens power and its energy dispersion.
We classify the resulting dispersion through a local Abbe number defined in the energy domain.

A radial Bessel singlet can retain a finite converging or diverging power while eliminating first-order axial chromatic aberration at a selected electron energy.
More importantly, spatially co-located doublets formed by combining the radial Bessel lens with either a magnetic objective lens or an azimuthally polarized Bessel lens can maintain a total electron-optical focal length of $4~\mathrm{mm}$ while canceling first-order chromatic and third-order spherical aberrations at the same time.
For the $1$- and $20$-keV cases examined here, the useful solutions occur mainly when the light propagates in the direction opposite to the electrons.

The finite-aperture calculations establish a practical trade-off among wavelength, required optical power, and higher-order spherical aberration.
At $1064~\mathrm{nm}$ the required power is compatible with demonstrated ultrashort-pulse lasers, although the higher-order spherical aberration restricts the usable electron-beam aperture.
At $10.6~\mathrm{\mu m}$ the broader field reduces this aberration, and the corrected doublets outperform the magnetic objective-lens model in a low-energy, large-energy-spread regime, but this wavelength demands a higher optical power and no source producing a comparable peak power is available at this wavelength.
The most realistic route to a proof-of-principle experiment is therefore a near-infrared single-pass geometry in which radially and azimuthally polarized annular laser pulses are focused by a common reflective objective lens.

Chromatic correction based on this mechanism is particularly useful for electron beams whose broad or discrete energy distributions must be preserved, such as radio-frequency-accelerated beams and beams prepared for attosecond bunching.
It may also stimulate the development of electron-optical systems that use several electron-beam energies in the way that optical microscopy uses several wavelengths.

The framework developed here provides a basis for compound electron-optical systems that incorporate a finite electron-lens separation, relay magnification, electrostatic elements of a different dispersion class, and more general vector optical fields.
The natural next step toward an achromatic optical corrector for an electron column is to extend the design freedom in this way and to search within it for the best trade-off against higher-order spherical aberration.

\section*{Acknowledgments}

We thank Professor Takumi Sannomiya for valuable advice and discussions throughout this work.
We are also deeply grateful to the late Genya Matsuoka, President of Lena Systems, Inc., for stimulating discussions and continuous encouragement.
This work was supported in part by JSPS KAKENHI Grant No. JP25K22000, JST FOREST Grant Nos. JPMJFR222F and JPMJFR223E, AMED under Grant No. JP24wm0625105, and NEDO (New Energy and Industrial Technology Development Organization) under Project No. JPNP14004.

\appendix

\section{Design formulas for scalar Bessel lenses}
\label{app:bessel_lens_formulas}

This appendix collects useful design formulas for scalar Bessel ponderomotive lenses.
The formulas are summarized from Refs.~\cite{Uesugi2022, Uesugi2023} and are presented here as a reference table.
The formulas in Ref.~\cite{Uesugi2022} were derived in the nonrelativistic limit.
Ref.~\cite{Uesugi2023} gives the corresponding formulation including the transverse-field relativistic scaling and discusses finite-thickness effects beyond the thin-lens approximation.

We consider a scalar Bessel ponderomotive potential of order $n$,
\begin{gather}
    U_n(r)=U_0 J_n^2(K_0 r),
    \label{eq:app_Bessel_potential}
\end{gather}
where 
\begin{gather}
    U_0 = \frac{h\,\alpha_\mathrm{fs}}{m\omega^2} I_0
    \label{eq:U0}
\end{gather}
is the nonrelativistic ponderomotive energy scale, $K_0=k\sin\theta_0$ is the transverse wavenumber, $\theta_0$ is the Bessel cone half-angle, $\alpha_\mathrm{fs}$ is the fine-structure constant, $\omega$ is the optical angular frequency, and $I_0$ is a time-averaged peak intensity scale used to normalize the field profile.
The electron kinetic energy is denoted by $T_0$ and the free-space electron momentum by $p_0$.

The formulas are obtained by expanding the canonical momentum near the optical axis as
\begin{gather}
    p(r,z)=p_0+p_2(z)r^2+p_4(z)r^4+p_6(z)r^6+\cdots .
    \label{eq:app_p_expansion}
\end{gather}
In the thin-lens approximation, the ray height is assumed to be unchanged while the electron passes through the optical potential.
The focusing power and the third- and fifth-order spherical-aberration coefficients are then given by the axial integrals
\begin{gather}
    \frac{1}{f}
    =
    -\frac{2}{p_0}
    \int_{-L/2}^{L/2}p_2(\zeta)\,d\zeta,
    \label{eq:app_thin_power_int}\\
    C_{S3}
    =
    -\frac{4a^4}{p_0}
    \int_{-L/2}^{L/2}p_4(\zeta)\,d\zeta,
    \label{eq:app_thin_Cs3_int}\\
    C_{S5}
    =
    -\frac{6a^6}{p_0}
    \int_{-L/2}^{L/2}p_6(\zeta)\,d\zeta,
    \label{eq:app_thin_Cs5_int}
\end{gather}
where $L$ is the effective interaction length, and $a$ is the object-to-aperture distance used in the spherical-aberration definition.

For an ideal Bessel field, the radial profile is independent of $z$.
In this case, $p_2$, $p_4$, and $p_6$ are constant over the effective interaction region, and the above integrals reduce to
\begin{gather}
    \frac{1}{f}
    =
    -\frac{2L}{p_0}p_2,
    \quad
    C_{S3}
    =
    -\frac{4a^4L}{p_0}p_4,
    \quad
    C_{S5}
    =
    -\frac{6a^6L}{p_0}p_6.
    \label{eq:app_thin_reduced}
\end{gather}
The parameter $L$ should therefore be understood as the effective axial interaction length.
For a physically generated Bessel-like field with a finite annular aperture, the axial envelope is finite; in that case, $U_0L$ in the table should be interpreted as the corresponding axial integral of the potential amplitude.

The formulas assume a weak ponderomotive potential, $|U_0|\ll T_0$, paraxial electron trajectories close to the optical axis, and a thin interaction region such that the ray height changes negligibly inside the optical field.
They are therefore design formulas for the paraxial focusing power and the lowest spherical-aberration coefficients, not a substitute for finite-thickness ray tracing when the optical interaction length becomes comparable to the electron-optical focal length or to other relevant propagation distances.

The expansion coefficients for Bessel ponderomotive lenses of orders $n=0$--3 are listed in Table~\ref{tab:app_Bessel_p_coefficients}.
The corresponding focusing powers and spherical-aberration coefficients are listed in Table~\ref{tab:app_Bessel_lens_properties}.

\begin{table}[t]
\caption{
Expansion coefficients of the canonical momentum for scalar Bessel ponderomotive lenses in the nonrelativistic thin-lens formulation.
}
\label{tab:app_Bessel_p_coefficients}
\begin{ruledtabular}
\begin{tabular}{ccccc}
 & $n=0$ & $n=1$ & $n=2$ & $n=3$ \\
\hline
$\displaystyle \frac{p_2}{p_0U_0/T_0}$
&
$\displaystyle \frac{K_0^2}{4}$
&
$\displaystyle -\frac{K_0^2}{8}$
&
$0$
&
$0$
\\[1.8ex]
$\displaystyle \frac{p_4}{p_0U_0/T_0}$
&
$\displaystyle -\frac{3K_0^4}{64}$
&
$\displaystyle \frac{K_0^4}{32}$
&
$\displaystyle -\frac{K_0^4}{128}$
&
$0$
\\[1.8ex]
$\displaystyle \frac{p_6}{p_0U_0/T_0}$
&
$\displaystyle \frac{5K_0^6}{1152}$
&
$\displaystyle -\frac{5K_0^6}{1536}$
&
$\displaystyle \frac{K_0^6}{768}$
&
$\displaystyle -\frac{K_0^6}{4608}$
\end{tabular}
\end{ruledtabular}
\end{table}

\begin{table}[t]
\caption{
Thin-lens properties of scalar Bessel ponderomotive lenses of orders $n=0$--3.
}
\label{tab:app_Bessel_lens_properties}
\begin{ruledtabular}
\begin{tabular}{ccccc}
 & $n=0$ & $n=1$ & $n=2$ & $n=3$ \\
\hline
$\displaystyle \frac{1/f}{U_0L/T_0}$
&
$\displaystyle -\frac{K_0^2}{2}$
&
$\displaystyle \frac{K_0^2}{4}$
&
$0$
&
$0$
\\[1.8ex]
$\displaystyle \frac{C_{S3}}{U_0K_0^4a^4L/T_0}$
&
$\displaystyle \frac{3}{16}$
&
$\displaystyle -\frac{1}{8}$
&
$\displaystyle \frac{1}{32}$
&
$0$
\\[1.8ex]
$\displaystyle \frac{C_{S5}}{U_0K_0^6a^6L/T_0}$
&
$\displaystyle -\frac{5}{192}$
&
$\displaystyle \frac{5}{256}$
&
$\displaystyle -\frac{1}{128}$
&
$\displaystyle \frac{1}{768}$
\end{tabular}
\end{ruledtabular}
\end{table}

The tables show that the $J_0^2$ lens has negative focusing power and acts as a diverging round lens, whereas the $J_1^2$ lens has positive focusing power and acts as a converging round lens.
Among the scalar Bessel lenses listed here, only the $J_1^2$ lens provides both positive focusing power and negative third-order spherical aberration.
The $J_2^2$ and $J_3^2$ lenses have no paraxial focusing power, but they can still generate higher-order spherical aberrations.

For a transversely polarized optical field, or for a scalar field treated with the relativistic transverse-field scaling, the nonrelativistic formulas are converted by the replacement
\begin{gather}
    T_0
    \longrightarrow
    \frac{T(1+\gamma)}{2}
    =
    \frac{(\gamma^2-1)mc^2}{2}
    =
    \frac{\beta^2\gamma^2mc^2}{2}.
    \label{eq:app_relativistic_replacement}
\end{gather}
This replacement reproduces the transverse-field relativistic scaling summarized in Ref.~\cite{Uesugi2023}.
It does not describe the longitudinally polarized relativistic correction discussed in the main text, for which the gauge-invariant treatment of the longitudinal field is required.

\section{Azimuthal-cone-parameter dependence of the azimuthal--radial design}
\label{app:saz_scan}

\begin{figure}
    \centering
    \includegraphics[width=\linewidth]{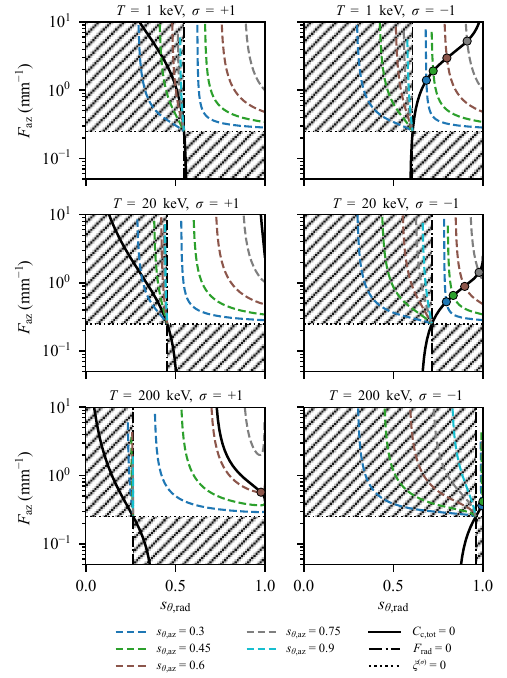}
    \caption{
    Design curves of the azimuthal--radial doublet for several azimuthal cone parameters $s_{\theta,\mathrm{az}}$ (colored dashed curves: $C_\mathrm{S3,tot}=0$) at electron energies of 1, 20, and 200 keV (rows) and propagation indices $\sigma=\pm1$ (columns), for $f_\mathrm{tot}=4$~mm and optical wavelengths of $1064~\mathrm{nm}$ for both Bessel fields.
    The solid $C_\mathrm{c,tot}=0$ curve is independent of $s_{\theta,\mathrm{az}}$.
    Dotted and dash-dotted lines mark $F_\mathrm{rad}=0$ and $\xi^{(\sigma)}=0$, hatching marks regions requiring negative radial excitation, and circles mark physical simultaneous solutions.
    }
    \label{fig:saz_scan}
\end{figure}

Figure~\ref{fig:saz_scan} generalizes the lower row of Fig.~\ref{fig:result_2} by varying the azimuthal cone parameter over $s_{\theta,\mathrm{az}}=0.3$--$0.9$ at electron energies of 1, 20, and 200 keV, with the remaining conditions unchanged.
Only the design curves and the physical-region boundaries are shown, and the vertical axis is restricted to a converging azimuthal reference lens, $F_\mathrm{az}>0$.
The $C_\mathrm{c,tot}=0$ curve is common to all $s_{\theta,\mathrm{az}}$ because the local Abbe number of the azimuthal Bessel lens is $\nu_\mathrm{az}=-1/2$ for any cone parameter [Eq.~(\ref{eq:nu_az})]; only the $C_\mathrm{S3,tot}=0$ curve moves, through the $K_\mathrm{az}^2$ dependence of the third-order kick coefficient $\Theta_{3,\mathrm{az}}$.
 
For $\sigma=-1$ at 1 and 20 keV, physical simultaneous solutions exist throughout $s_{\theta,\mathrm{az}}\simeq0.3$--$0.75$.
Increasing $s_{\theta,\mathrm{az}}$ displaces the solution toward larger $s_{\theta,\mathrm{rad}}$ and larger $F_\mathrm{az}$, that is, toward a higher radial numerical aperture and a stronger internal cancellation between the component powers; at $s_{\theta,\mathrm{az}}=0.9$ the solution leaves the accessible range $s_{\theta,\mathrm{rad}}<1$.
A smaller $s_{\theta,\mathrm{az}}$ is therefore favorable in both respects, at the cost of a larger azimuthal excitation, since $(U_0l)_\mathrm{az}\propto s_{\theta,\mathrm{az}}^{-2}$ at fixed $F_\mathrm{az}$.
 
Raising the electron energy compresses the accessible $\sigma=-1$ solutions toward $s_{\theta,\mathrm{rad}}\to1$: at 200 keV, solutions with a converging azimuthal reference remain only for $s_{\theta,\mathrm{az}}\lesssim0.45$, with $s_{\theta,\mathrm{rad}}\approx0.99$.
At this energy, a $\sigma=+1$ solution appears for $s_{\theta,\mathrm{az}}\approx0.5$--$0.6$, likewise at $s_{\theta,\mathrm{rad}}\gtrsim0.98$.

\section{Transverse-aberration evaluation}
\label{app:transverse_aberration_evaluation}

The transverse-aberration plots use an electron-energy interval
\begin{equation}
    T_\pm
    =
    T_0
    \pm
    \frac{\Delta T}{2}.
    \label{eq:appendix_energy_interval}
\end{equation}

The diffraction-limited disk radius is
\begin{equation}
    \Delta r_\mathrm{d}
    =
    \frac{
    0.61\lambda_\mathrm{e}(T_0)
    }{
    \alpha
    },
    \label{eq:appendix_diffraction}
\end{equation}
where $\lambda_\mathrm{e}$ is the relativistic electron wavelength and $\alpha$ is the convergence semi-angle.

When the spherical aberration is represented through fifth order, the signed focal-plane displacement is
\begin{equation}
    \delta r_\mathrm{s}^{(3+5)}
    =
    C_\mathrm{S3}\alpha^3
    +
    C_\mathrm{S5}\alpha^5,
\end{equation}
and the corresponding disk radius is
\begin{equation}
    \Delta r_\mathrm{s}^{(3+5)}
    =
    \left|
    C_\mathrm{S3}\alpha^3
    +
    C_\mathrm{S5}\alpha^5
    \right|.
    \label{eq:appendix_spherical}
\end{equation}

For a lens represented by a first-order chromatic coefficient
$C_\mathrm{c}$, the disk radius over the full energy width
$\Delta T$ is
\begin{equation}
    \Delta r_\mathrm{c}
    =
    \frac12
    \left|
    C_\mathrm{c}
    \right|
    \alpha
    \frac{\Delta T}{T_0}.
    \label{eq:appendix_chromatic}
\end{equation}

The total transverse radius is evaluated by the quadrature sum
\begin{equation}
    \Delta r_\mathrm{tot}
    =
    \sqrt{
    \Delta r_\mathrm{d}^2
    +
    \Delta r_\mathrm{s}^2
    +
    \Delta r_\mathrm{c}^2
    }.
    \label{eq:appendix_total}
\end{equation}

A direct minimization of Eq.~(\ref{eq:appendix_total}) can select a narrow cancellation between the third- and fifth-order spherical terms.
For the third- plus fifth-order model, the optimum semi-angle is therefore selected by minimizing
\begin{equation}
    \sqrt{
    \Delta r_\mathrm{d}^2
    +
    \Delta r_\mathrm{c}^2
    +
    \left(
    \left|
    C_\mathrm{S3}
    \right|
    \alpha^3
    +
    \left|
    C_\mathrm{S5}
    \right|
    \alpha^5
    \right)^2
    }.
    \label{eq:appendix_robust_35}
\end{equation}
The curves shown in the figures nevertheless use the signed sum in Eq.~(\ref{eq:appendix_spherical}).

For a non-expanded Bessel calculation, isolated zeros and oscillatory minima are suppressed by replacing the spherical term in the optimization criterion by
\begin{equation}
    \max_{0\leq\alpha'\leq\alpha}
    \Delta r_\mathrm{s}^{(\mathrm{full})}
    \left(
    \alpha'
    \right).
    \label{eq:appendix_full_envelope}
\end{equation}
The plotted full-Bessel curves themselves use the unmodified
$\Delta r_\mathrm{s}^{(\mathrm{full})}$ and
$\Delta r_\mathrm{tot}^{(\mathrm{full})}$.

\section{Power estimate for a finite Bessel-like field}
\label{app:bessel_power_estimate}

We estimate the optical power required to realize a finite Bessel-like field whose ideal transverse intensity is written as an incoherent sum of \(J_0^2\) and \(J_1^2\) Bessel profiles.
The angle \(\theta_0\) denotes the design-center Bessel cone half-angle, and
\begin{equation}
    s_0=\sin\theta_0,\qquad
    c_0=\cos\theta_0 .
\end{equation}
The ideal transverse intensity coefficient is defined by
\begin{equation}
    I(r,z)
    =
    I_0 |g(z)|^2
    \left[
        a_0 J_0^2(K r)
        +
        a_1 J_1^2(K r)
    \right],
\end{equation}
where \(I_0\) is the overall intensity scale, and \(a_0\) and \(a_1\) are dimensionless weights for the incoherent \(J_0^2\) and \(J_1^2\) Bessel components.
They are normalized as \(a_0+a_1=1\).

The function $g(z)$ is a dimensionless longitudinal envelope introduced to describe the finite axial extent of the Bessel-like field, and is normalized as $g(0)=1$.
The corresponding effective interaction length is defined by
\begin{equation}
    l_{\rm eff}
    =
    \int_{-\infty}^{\infty}|g(z)|^2\,dz .
    \label{eq:leff_def}
\end{equation}

We model the finite Bessel-like field by a rectangular annular angular spectrum.
The annulus is centered at \(s=s_0\) and has full width \(\Delta s\), i.e.,
\begin{equation}
    s_0-\Delta s/2
    <
    s
    <
    s_0+\Delta s/2 .
    \label{eq:rect_annulus}
\end{equation}
Here $s=\sin\theta$ is the angular-spectrum coordinate, and the field amplitude is taken to be constant inside the interval in Eq.~(\ref{eq:rect_annulus}).
For a narrow annulus, the longitudinal wavenumber is linearized around $s=s_0$ as
\begin{equation}
    k_z(s)
    =
    k\sqrt{1-s^2}
    \simeq
    k c_0
    -
    \frac{k s_0}{c_0}(s-s_0).
    \label{eq:kz_linear}
\end{equation}
With Eq.~(\ref{eq:kz_linear}), the normalized longitudinal envelope for the rectangular annulus is
\begin{equation}
    g(z)
    =
    e^{ikc_0z}
    \mathrm{sinc}\!\left(
        \frac{k s_0\Delta s}{2c_0}z
    \right),
    \label{eq:g_rect}
\end{equation}
where \(\mathrm{sinc}\,x\equiv \sin x/x\).
Substitution of Eq.~(\ref{eq:g_rect}) into Eq.~(\ref{eq:leff_def}) gives
\begin{equation}
    l_{\rm eff}
    =
    \frac{\lambda c_0}{s_0\Delta s}.
    \label{eq:leff_rect}
\end{equation}

We now relate the intensity scale $I_0$ to the optical power $P$ crossing a plane normal to the optical axis.
The quantity $I_0$ is defined as the field-intensity scale relevant to the ponderomotive interaction, i.e., a quantity proportional to $|E|^2$.
It is therefore not, by itself, the axial power flux.
In the scalar conical-wave model used here, each constituent wave propagates at the cone half-angle $\theta_0$, so only the fraction $c_0=\cos\theta_0$ of its energy flux crosses a transverse plane.

For a single Bessel component truncated at transverse radius $R$, the corresponding axial optical power is approximated as
\begin{equation}
    P
    =
    c_0\,2\pi a_m I_0
    \int_0^R J_m^2(Kr)\,r\,dr
    \simeq
    c_0 a_m I_0 \frac{2R}{K}.
    \label{eq:Pm_R}
\end{equation}
The last approximation assumes $KR\gg1$ and is independent of $m$ to leading order for $m=0,1$.
Therefore, the same optical-power estimate applies to the $J_0^2$ and $J_1^2$ components when their weights are defined through the common scale $I_0$.

The transverse truncation radius $R$ can be expressed as a two-sided Bessel-zone length,
\begin{equation}
    l=\frac{2R}{\tan\theta_0}.
    \label{eq:l_R_relation}
\end{equation}
Using $K=ks_0$ and $\tan\theta_0=s_0/c_0$, Eqs.~(\ref{eq:Pm_R})--(\ref{eq:l_R_relation}) give
\begin{equation}
    P_m
    \simeq
    a_m\frac{I_0l}{k}
    =
    a_m\frac{\lambda}{2\pi}I_0l .
    \label{eq:Pm_l}
\end{equation}
Thus, after identifying $l$ with the effective length $l_{\rm eff}$ defined in Eq.~(\ref{eq:leff_def}), the total optical power of the incoherent sum is
\begin{equation}
    P_\mathrm{tot}
    =
    (a_0+a_1)
    \frac{\lambda}{2\pi}I_0l_{\rm eff}.
    \label{eq:P_sum}
\end{equation}

Using $l=l_{\rm eff}$ in Eq.~(\ref{eq:l_R_relation}), together with Eq.~(\ref{eq:leff_rect}), gives
\begin{equation}
    KR
    =
    \frac{\pi s_0}{\Delta s}.
    \label{eq:KR_condition}
\end{equation}
Hence the condition $KR\gg1$ is equivalent, in this rectangular-annulus model, to a sufficiently narrow annulus, $\Delta s/s_0\ll\pi$.

Finally, suppose that the interaction calculation specifies the integrated ponderomotive parameter
\begin{equation}
    Q
    =
    U_0 l
    \simeq
    \mu I_0 l_{\rm eff},
    \label{eq:Q_def}
\end{equation}
where $\mu = h \alpha_{\mathrm{fs}} / m\omega^2$.
Substituting Eq.~(\ref{eq:Q_def}) into Eq.~(\ref{eq:P_sum}) with $a_0+a_1=1$ gives
\begin{equation}
    P_{\mathrm{tot}}
    =
    \frac{\lambda Q}{2\pi\mu}.
    \label{eq:P_from_Q}
\end{equation}

This shows that, within the rectangular-annulus model, the required optical power for a specified integrated interaction strength $U_0 l$ is determined by the optical wavelength and the interaction coefficient, but not by the cone angle $\theta_0$ or the annular width $\Delta s$.
The cancellation occurs because $s_0$, $c_0$, and $\Delta s$ enter both the effective length $l_{\rm eff}$ and the local intensity scale $I_0$.

This order estimate rests on a scalar narrow-annulus model, and two of its assumptions warrant separate comment.
First, the scalar treatment itself holds only well below the high-numerical-aperture regime; at high NA, vectorial effects and pupil apodization introduce additional corrections, depending on the cone angle and the annular profile, that are not captured here.
Second, the narrow-annulus condition, $\Delta s/s_0\ll\pi$, is itself an idealization. Finite annular pupils producing tightly focused radially polarized Bessel-like fields have nonetheless been implemented experimentally \cite{Tsuru2024}, indicating that annuli narrow enough for the present estimate to be a useful guide are experimentally accessible.

\end{document}